\documentclass[%
 aip,.pdf
 amsmath,amssymb,
 reprint,%
nolinenumbers
]{revtex4-2}
\usepackage{ulem}
\usepackage{amsmath}
\usepackage{amssymb}
\usepackage{graphicx}
\usepackage{dcolumn}
\usepackage{bm}
\usepackage{color}
\usepackage{xcolor}
\usepackage[utf8]{inputenc}
\usepackage[T1]{fontenc}
\usepackage{mathptmx}
\usepackage{etoolbox}
\usepackage{tabularx}
\usepackage{lineno}
\usepackage{soul}
\usepackage{tocloft}

\cftpagenumbersoff{figure}

\makeatletter
\def\@email#1#2{%
 \endgroup
 \patchcmd{\titleblock@produce}
  {\frontmatter@RRAPformat}
  {\frontmatter@RRAPformat{\produce@RRAP{*#1\href{mailto:#2}{#2}}}\frontmatter@RRAPformat}
  {}{}
}%
\makeatother
\begin{document}

\preprint{AIP/123-QED}

\title[]{Drastic enhancement of the superconducting temperature in type-II Weyl semimetal candidate MoTe$_2$ via biaxial strain}
\author{King Yau Yip}
\author{Siu Tung Lam}
\author{Kai Ham Yu}
\author{Wing Shing Chow}
\author{Jiayu Zeng}
\affiliation{Department of Physics, The Chinese University of Hong Kong, Shatin, Hong Kong, China} 
\author{Kwing To Lai$^*$}
\affiliation{Department of Physics, The Chinese University of Hong Kong, Shatin, Hong Kong, China} 
\affiliation{Shenzhen Research Institute, The Chinese University of Hong Kong, Shatin, Hong Kong, China}
\affiliation{Faculty of Science, The University of Hong Kong, Pokfulam Road, Hong Kong, China} 
\author{Swee K. Goh$^*$}
\email{ktlai@phy.cuhk.edu.hk and skgoh@cuhk.edu.hk}
\affiliation{Department of Physics, The Chinese University of Hong Kong, Shatin, Hong Kong, China}

\date{\today}

\begin{abstract}
Type-II Weyl semimetal candidate MoTe$_2$, which superconducts at $T_c\sim$0.1~K, is one of the promising candidates for realizing topological superconductivity. However, the exceedingly low $T_c$ is associated with a small upper critical field ($H_{c2}$), implying a fragile superconducting phase that only exists on a small region of the $H$-$T$ phase diagram. Here, we describe a simple and versatile approach based on the differential thermal expansion between dissimilar materials to subject a thin single crystalline MoTe$_2$ to biaxial strain. With this approach, we successfully enhance the $T_c$ of MoTe$_2$ by five-fold and consequently expand the superconducting region on the $H$-$T$ phase diagram significantly. To demonstrate the relative ease of studying the superconductivity in the biaxially strained MoTe$_2$, we further present the magnetotransport data, enabling the study of the temperature-dependent $H_{c2}$ and the anisotropy of the superconducting state which would otherwise be difficult to obtain in a free-standing MoTe$_2$. Our work shows that biaxial strain is an effective knob to tune the electronic properties of MoTe$_2$. Due to the simplicity of our methodology to apply biaxial strain, we anticipate its direct applicability to a wider class of quantum materials. \\ 

\end{abstract}

\maketitle
The combination of topological band structure and superconductivity offers a viable route to search for topological superconductivity, where Majorana fermions can be used for topological quantum computation \cite{Sato2017,Li2019aqt}. 
Type-II Weyl semimetal candidate MoTe$_2$, which is also a superconductor, is one of the promising candidates for hosting topological superconductivity \cite{Soluyanov2015,Deng2016,Huang2016,Jiang2017nc}. This belief is further strengthened after the discovery of an edge supercurrent in MoTe$_2$~\cite{Wang2020}. 

MoTe$_2$ has a layered crystal structure with Mo-Te layers stacked along the $c$-axis. Upon cooling, MoTe$_2$ undergoes a first-order structural transition at $T_s \sim$ 250~K, changing from a centrosymmetric monoclinic $1T'$ phase (space group: $P2_1/m$) to a noncentrosymmetric orthorhombic $T_d$ phase (space group: $Pmn2_1$). With further cooling to low temperatures, a superconducting phase transition at $T_c \sim$ 0.1~K can be observed in a free-standing sample at ambient pressure~\cite{Qi2016,Hu2019}. Despite the keen interest in the superconducting state of MoTe$_2$, its low $T_c$ poses a significant experimental challenge. Furthermore, such a low $T_c$ also brings along small upper critical fields ($H_{c2}$) along all field directions, impeding a powerful examination of the superconducting state with a magnetic field~\cite{Hu2019,Luo2020}. 

Various efforts have been devoted to enhancing the $T_c$ of MoTe$_2$. When MoTe$_2$ is subjected to hydrostatic pressure, its $T_c$ can indeed be dramatically enhanced -- $T_c$ increases by 30-fold ($\sim$4~K) at $\sim$15~kbar. Furthermore, the enhancement of $T_c$ by hydrostatic pressure is accompanied by the suppression of a structural transition temperature, resulting in an intricate phase diagram showing the competition between structural and superconducting transitions  \cite{Qi2016,Takahashi2017,Heikes2018,Lee2018,Guguchia2017,Hu2019}. A similar $T_c$ enhancement can also be observed via chemical substitutions \cite{Takahashi2017,Chen2016,Cho2017,Mandal2018,Dahal2020,Mandal2021}.

Although hydrostatic pressure and chemical substitutions have successfully enhanced $T_c$, the former requires specialized equipment and knowledge while the latter unavoidably introduces disorder. Moreover, in a high-pressure experiment, the sample space is enclosed by high-strength materials, limiting the accessibility of most experimental probes. Thus, a simpler tuning method with an open architecture is welcomed. 
Recently, biaxial strain emerges as an innovative tuning technique for studying superconductivity~\cite{Bohmer2017,Nakajima2021}. Different from uniaxial strain, biaxial strain applies simultaneous stresses to a sample in two orthogonal directions. The sample can then be tuned with in-plane symmetry preserved, offering a new way to discern the interplay between different quantum phases and superconductivity. Among several existing methods to achieve biaxial strain, a relatively convenient way is to attach the sample onto a substrate with a thermal expansion/contraction largely different from that of the sample at low temperatures, so that biaxial strain is induced on the sample when the whole device is cooled. This elegant method has been employed to study iron-based superconductors~\cite{Bohmer2017,Nakajima2021}. In this manuscript, we apply biaxial strain to MoTe$_2$ by attaching a thin flake of MoTe$_2$ onto a Stycast substrate to tune its superconductivity. We successfully enhance the $T_c$ by five-fold and enlarge the superconducting region in the $H$-$T$ phase diagram, enabling a detailed analysis of $H_{c2}$. To demonstrate the quality of our device, we further measure quantum oscillations in resistance up to 14~T.

\begin{figure}[!t]\centering
      \resizebox{8cm}{!}{
              \includegraphics{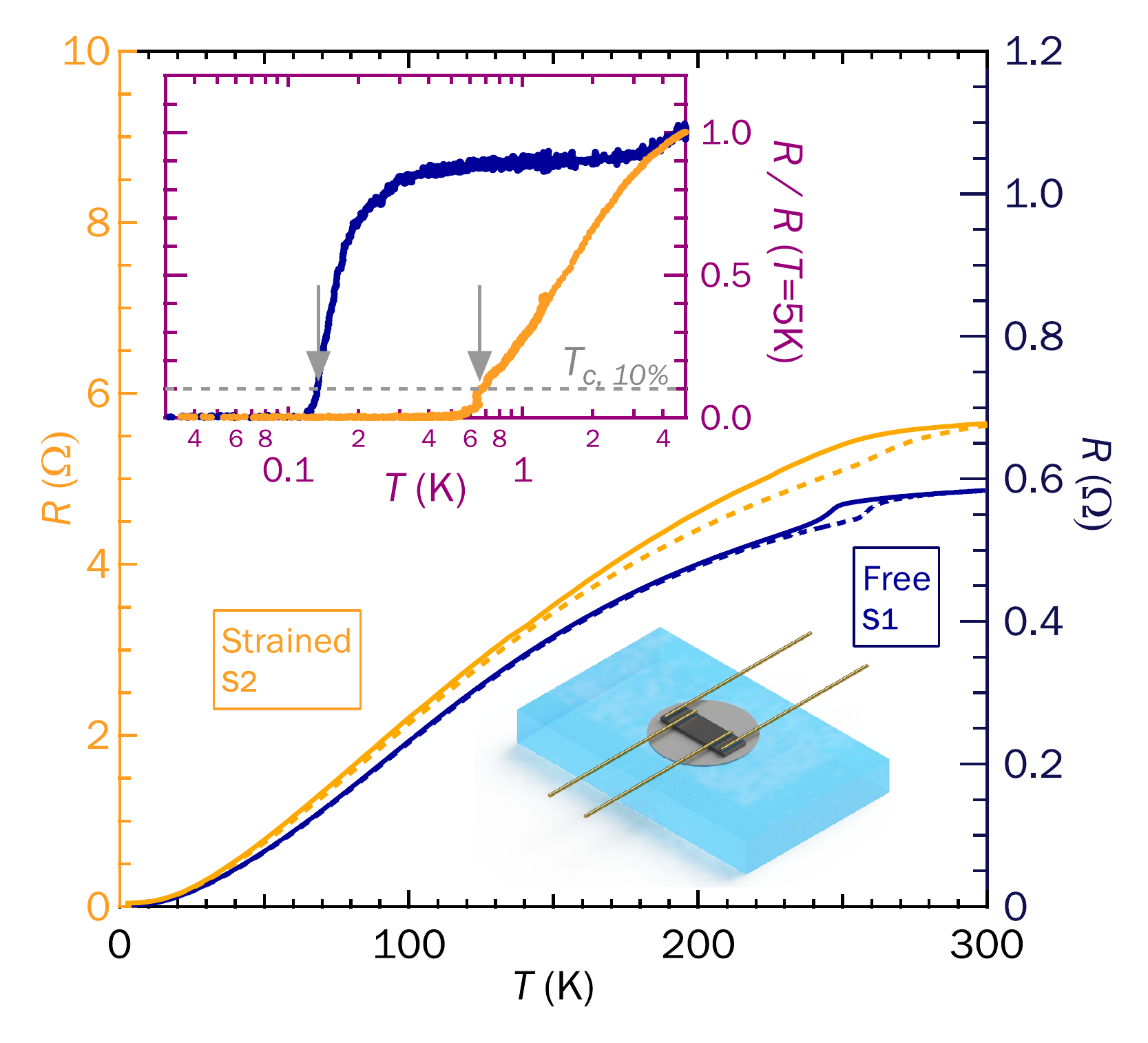}}                				
              \caption{\label{fig1} Temperature dependence of electrical resistance $R(T)$ of free-standing (S1) and biaxially strained (S2) MoTe$_2$ at zero magnetic field. The cool-down (warm-up) curves are plotted as solid (dashed) curves. 
              The top inset shows normalized $R(T)$ curves for S1 (navy) and S2 (orange), displaying their superconducting transitions. The Horizontal dashed grey line indicates the `$10\%$ criterion' for the definition of $T_c$ (see Supplementary Material for details). The bottom inset displays the schematic setup of biaxially strained MoTe$_2$ in a four-probe configuration with CN adhesive under the sample (circular) and the Stycast 1266 substrate underneath (rectangular).}
\end{figure}


Single crystals of MoTe$_2$ were prepared by the self-flux method as described elsewhere \cite{Hu2020}. Two single crystals (S1 and S2) grown from the same batch were chosen in this work. The dimensions for S1 and S2 are 1070~$\times$~487~$\times$~11.3~$\mu$m$^3$ and 937~$\times$~128~$\times$~7.2~$\mu$m$^3$ ($L\times W \times H$), respectively. S1 was used as a `free-standing' reference while S2 is chosen for the exertion of biaxial strain. To strain the sample, S2 was attached onto the substrate made of pre-cured Stycast 1266 A/B epoxy (Henkel Loctite). The thickness of the substrate is 3.08~mm. Cyanoacrylate (CN) adhesives (Tokyo Measuring Instruments Lab Co., Ltd.), which are mechanically stable at cryogenic temperatures, were used as the strain-transmitting medium. A schematic setup was shown in the bottom inset of Fig.~\ref{fig1}. A standard four-probe method was used to measure temperature-dependent resistance in a Bluefors dilution refrigerator down to 20~mK. Magnetic field was applied with a 3~T/5~T (y/z-axes) superconducting vector magnet for $H_{c2}$ studies. Quantum oscillations in resistance were additionally measured on S2 at 2~K up to 14~T in Physical Property Measurement System (PPMS) by Quantum Design.


Figure~\ref{fig1} shows the temperature dependence of the zero-field electrical resistance $R(T)$ for both the free-standing (S1) and biaxially strained MoTe$_2$ (S2). In this experimental setup (lower inset of Fig.~\ref{fig1}), biaxial strain induced on S2 is directly proportional to the difference between the in-plane thermal expansion ($\Delta L/L$) of the free-standing MoTe$_2$ and the substrate. Based on the fact that $\Delta L/L$ of MoTe$_2$ has a relatively weak temperature dependence~\cite{Heikes2018}, the biaxial strain experienced by S2 is dominated by the temperature dependence of $\Delta L/L$ of Stycast 1266, which can be found in Ref.~[\onlinecite{Swift1979}]. Based on these considerations, we can estimate the magnitude of biaxial strain on S2. 

The $R(T)$ curves in Fig.~\ref{fig1} exhibit an anomaly at around $T_s\approx 250$~K accompanied by a pronounced thermal hysteresis, which is consistent with a first-order structural phase transition~\cite{Qi2016,Hu2019,Hu2020}. The values of $T_s$ for S1 and S2 are similar because the effect of biaxial strain is rather small at high temperatures.  The smearing of the structural phase transition in S2 is attributed to the continuously varying biaxial strain over the high-temperature range. 
At the 0~K limit, thermal expansion must stop because of the third law of thermodynamics. Therefore, the biaxial strain on S2 saturates at the lowest temperatures (equivalent to a compressive strain $\varepsilon_{biaxial}\sim0.6\%~$ below 3~K). This provides a simple platform to investigate the effect of biaxial strain on the superconductivity of MoTe$_2$. The top inset of Fig.~\ref{fig1} shows clear superconducting transitions where the resistances of S1 and S2 drop abruptly to zero. The values of the superconducting critical temperature ($T_c$), defined at which the resistance drops to 10~\% of the resistance at the normal state, are found to be $\sim 130$~mK and $\sim 640$~mK for S1 and S2, respectively. Thus, in the presence of a biaxial strain, $T_c$ of S2 exhibits a five-fold enhancement compared with the free-standing case (S1). 

\begin{figure}[!t]\centering
       \resizebox{7.5cm}{!}{
              \includegraphics{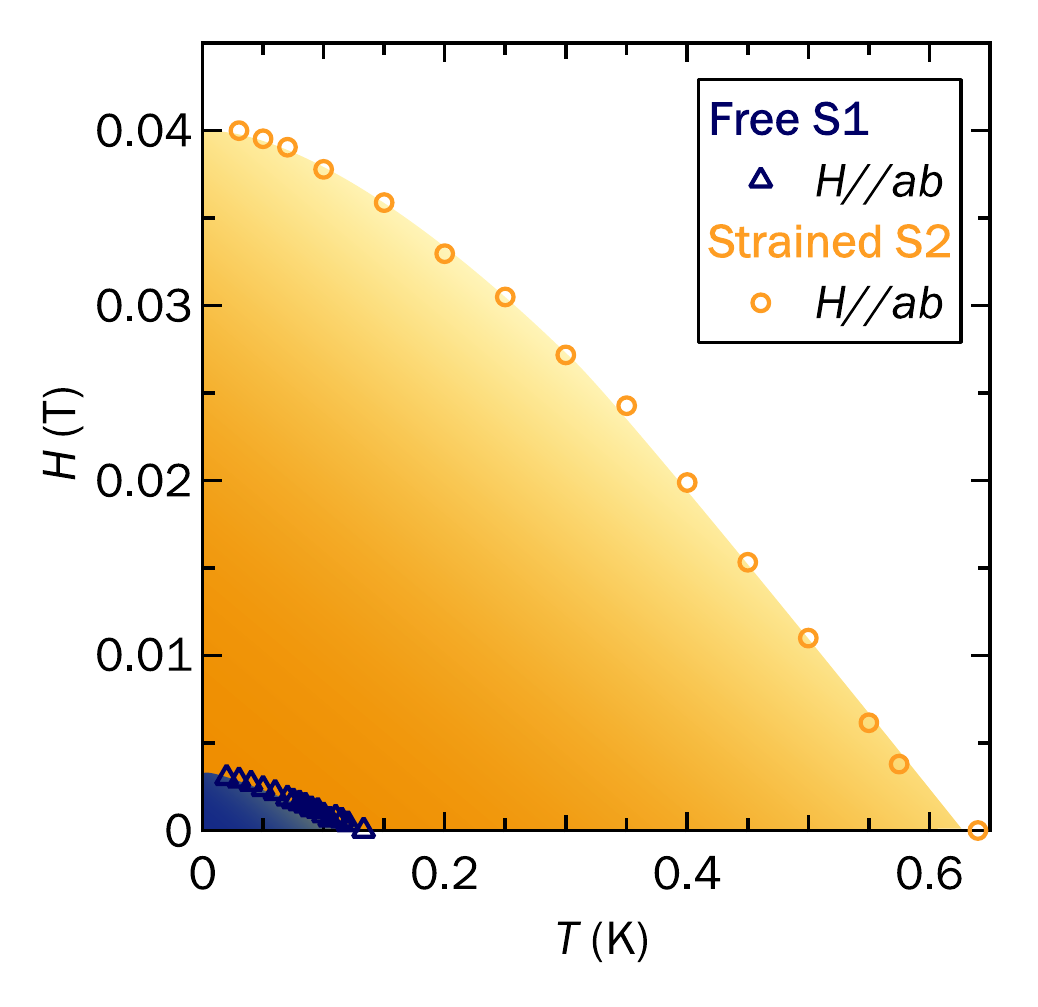}}  
              
              \caption{\label{fig2} Temperature dependence of upper critical field $H_{c2}(T)$ of free standing (S1) and biaxially strained (S2) MoTe$_2$ with the magnetic field $H$ along $ab$-plane ($H\parallel ab$). The superconducting regions enclosed by $H_{c2}(T)$ data (open symbols) for S1 and S2 are shaded as a visual guide. In addition to the five-fold enhancement of $T_c$ at zero field, the $H_{c2}$ has shown a ten-fold enhancement near absolute zero.
               }
\end{figure}

Figure~\ref{fig2} displays the temperature dependence of the in-plane upper critical field ($H^{\parallel ab}_{c2}$) for S1 and S2. The raw data from which the upper critical fields are determined are shown in Supplementary Material. For consistency, the same `10\% criterion' is used to determine the $H_{c2}$ under different conditions throughout this manuscript. The contrasting behaviour between S1 and S2 immediately demonstrates the expansion of the superconducting region on the $H$-$T$ phase diagram when biaxial strain is applied, consistent with the strengthening of the superconducting phase.

\begin{figure}[!t]\centering
       \resizebox{8.2cm}{!}{
              \includegraphics{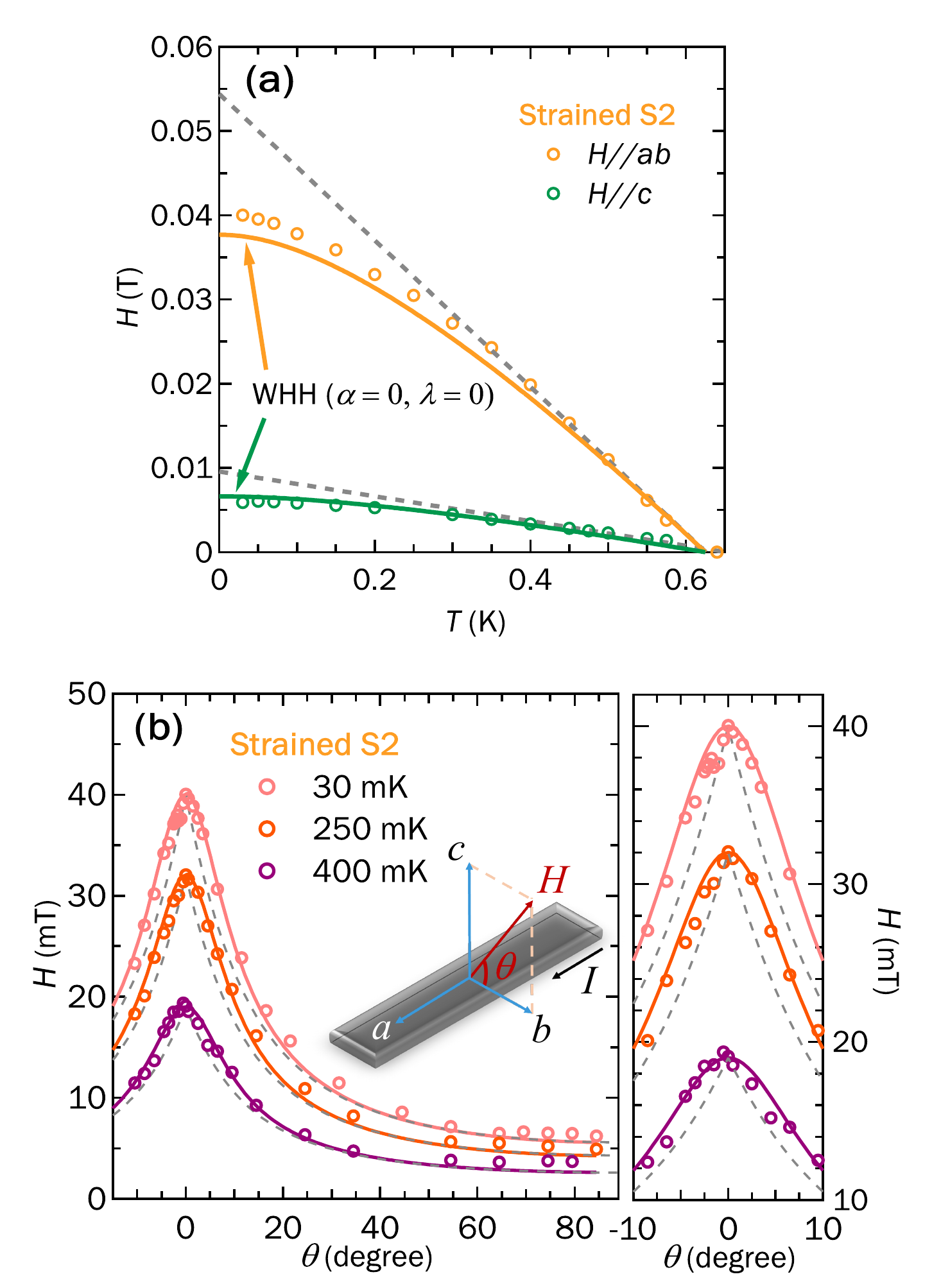}}                				
              \caption{\label{fig3} (a) Temperature dependence of upper critical field $H_{c2}(T)$ of biaxially strained MoTe$_2$ (S2) with $H\parallel ab$ (orange) and $H\parallel c$ (green). Dashed grey lines are the linear fits of $H_{c2}(T)$ near $T_c$. Solid curves represent the simulated $H_{c2}(T)$ curves with the single-band WHH model.
              (b) Angular dependence of upper critical field $H_{c2}(\theta)$ of S2. Left panel displays $H_{c2}(\theta)$ measured at 30~mK, 250~mK, and 400~mK. Here the solid (dashed) curves represent the 3D GL model (2D TK model). The inset illustrates the arrangement for $H_{c2}(\theta)$ measurements, showing the field angle $\theta$ and the crystallographic axes of the sample. The current direction is always perpendicular to the field direction. The right panel focuses on the data near $\theta=0^{\circ}$ ($H\parallel ab$).} 
\end{figure}

The enhanced $H_{c2}$ offers the opportunity to investigate the pair-breaking mechanism and the anisotropy of the superconducting state. Figure~\ref{fig3}(a) shows both the in-plane and the out-of-plane upper critical fields (denoted $H^{\parallel ab}_{c2}$ and $H^{\parallel c}_{c2}$, respectively) on the $H$-$T$ plane for S2. The collection of $H^{\parallel c}_{c2}$ would be prohibitively difficult for S1, as the superconducting state is extremely fragile for $H\parallel c$. For S2, both $H_{c2}(T)$ curves show a linear dependence near $T_c$, a characteristic of the pair-breaking mechanism by the orbital effect. In the Werthamer-Helfand-Hohenberg (WHH) model for a type-II superconductor in the dirty limit, the orbital-limited field at $T$= 0~K is given by~\cite{Werthamer1966}
\begin{equation} \label{orbit}
\left.H_{c2}^{orb}(0)=-0.693\times T_c\dfrac{dH_{c2}}{dT}\right\vert_{T=T_c}.
\end{equation}
The initial slopes $(dH_{c2}/dT)_{T=T_c}$, obtained by the linear fits near $T_c$ (dashed lines in Fig.~\ref{fig3}(a)), for $H\parallel ab$ and $H\parallel c$ are found to be $-0.087$~T/K and $-0.015$~T/K, respectively. Using this information, we simulate $H_{c2}(T)$ curves with the single-band WHH model \cite{Werthamer1966} with the Maki parameter $\alpha=0$ and spin-orbit coupling $\lambda=0$ for both field directions (solid curves), as illustrated in Fig.~\ref{fig3}(a). The experimental data (solid circles) agree quite well with the corresponding WHH curves, except for a small enhancement observed in the data of $H\parallel ab$ at low temperatures. Therefore, the orbital pair-breaking effect is dominant in both field directions.

The Ginzburg-Landau coherence lengths of the Cooper pairs in the $ab$-plane ($\xi_{ab}$) and the $c$-axis ($\xi_{c}$) of S2 can be extracted from the $H_{c2}$ data in Fig.~\ref{fig3}(a) through the following equations~\cite{tinkham2004}: 
\begin{equation} \label{xi_ab}
H^{\parallel ab}_{c2} = \frac{\phi_0}{2\pi\xi_{ab}\xi_{c}},    
\end{equation}
\begin{equation} \label{xi_c}
H^{\parallel c}_{c2} = \frac{\phi_0}{2\pi{\xi_{ab}}^2}, 
\end{equation}
where $\phi_0$ is the flux quantum. The values of $\xi_{ab}$ and $\xi_{c}$ at 30~mK are calculated to be 236~nm and 35~nm, respectively. Using the definition of an anisotropy factor $\gamma=\xi_{ab}/\xi_{c}$, the anisotropy of S2 is found to be moderate ($\gamma\sim$7), from which it is, however, difficult to infer the dimensionality of superconductivity. Furthermore, although a large $\gamma$ (i.e., a large anisotropy in superconducting properties) is expected in quasi-2D layered materials, exceptions have been reported in, e.g., our previous studies on some of the transition metal dichalcogenides (TMD) \cite{Chan2017,Hu2019}. Therefore, using $\gamma$ alone is not sufficient to determine the effect of biaxial strain on the dimensionality of superconductivity in MoTe$_2$. 

\begin{figure}[!t]\centering
       \resizebox{8cm}{!}{
              \includegraphics{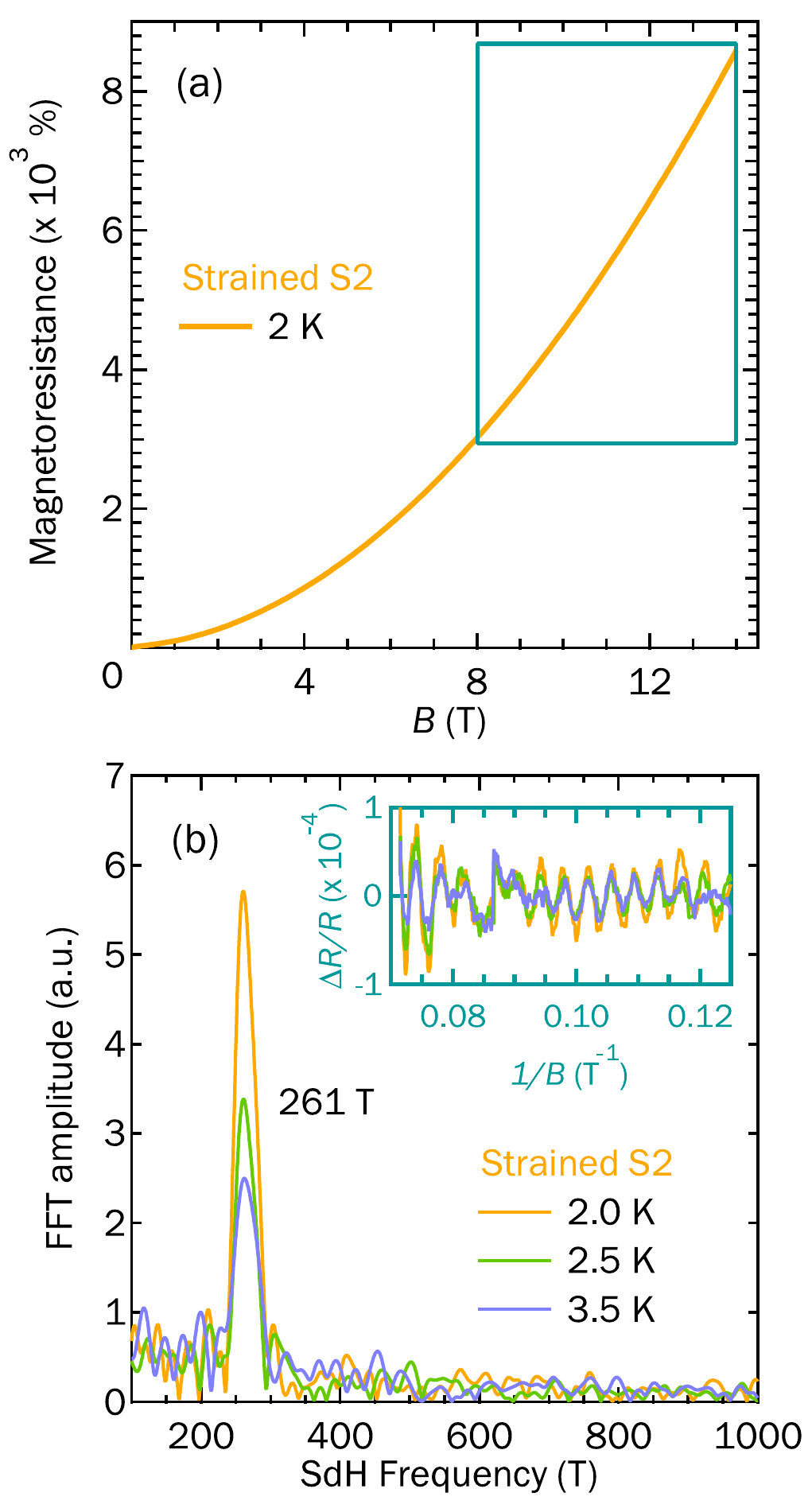}}                				
              \caption{\label{fig4} (a) Magnetoresistance of biaxially strained MoTe$_2$ (S2) from 0~T to 14~T measured at 2~K. The rectangle encloses the data used for extracting quantum oscillations shown in the inset of (b). (b) Fast Fourier transform spectra of the SdH oscillations shown in the inset. All the spectra exhibit a well-defined peak with an oscillation frequency of 261~T, whose amplitude decreases with increasing temperatures. Inset: Shubnikov-de Haas (SdH) oscillations of S2 at 2~K, 2.5~K, and 3~K.}
\end{figure}

To directly probe the superconducting dimensionality, we report the angular dependence of upper critical field $H_{c2}(\theta)$ for S2, as displayed in Fig.~\ref{fig3}(b). All data (open symbols) presented were obtained by field sweeps at fixed $\theta$ and $T$. The following two models are commonly used to analyze $H_{c2}(\theta)$: the anisotropic mass Ginzburg-Landau model (3D GL model) and the Tinkham model (2D TK model), which describe the existence of 3D and 2D superconductivity, respectively~\cite{Tinkham1963}. Their functional forms at a constant temperature are given, respectively, by Eq.~\eqref{GL} and Eq.~\eqref{Tinkham}:

\begin{equation} \label{GL} \centering
\left[\dfrac{H_{c2}(\theta)\mathrm{sin}(\theta)}{H_{c2}(90^{\circ})}\right]^2+\left[\dfrac{H_{c2}(\theta)\mathrm{cos}(\theta)}{H_{c2}(0^{\circ})}\right]^2=1
\end{equation}

\begin{equation} \label{Tinkham} \centering
\left|\dfrac{H_{c2}(\theta)\mathrm{sin}(\theta)}{H_{c2}(90^{\circ})}\right|+\left[\dfrac{H_{c2}(\theta)\mathrm{cos}(\theta)}{H_{c2}(0^{\circ})}\right]^2=1
\end{equation}

The key difference between these two models is the behaviour of $H_{c2}(\theta)$ when $\theta \rightarrow0^{\circ}$ ($H\parallel ab$): while the 3D GL model shows a smooth maximum, the 2D TK model shows a cusp-like behaviour due to the term with an absolute sign in Eq.~\eqref{Tinkham}. To reveal the dimensionality of superconductivity, we carefully analyze the data of S2 with both models. We find that $H_{c2}(\theta)$ of S2 can be well described by the 3D GL model. Thus, despite the layered structure, MoTe$_2$ under biaxial strain is a 3D superconductor with a moderate anisotropy factor. This finding is reminiscent of $H_{c2}(\theta)$ of WTe$_2$ under pressure~\cite{Chan2017}, which also follows the 3D GL model, although the anisotropy factor is a lot smaller there ($\gamma\approx$1.7).

Having demonstrated the ability to have a deeper investigation of the superconducting state when MoTe$_2$ is subjected to a modest biaxial strain, we briefly discuss the prospect of Fermiology studies with the same device. This is an important aspect because, at the microscopic level, biaxial strain modifies the electronic structure, affecting the Fermi surfaces of metallic systems. With our device, the most straightforward method is to search for magnetic quantum oscillations in magnetoresistance, in which the quantum oscillation frequency directly gives us the area of the extremal orbit on the Fermi surface perpendicular to the magnetic field direction.  Figure~\ref{fig4}(a) shows the magnetoresistance of S2 at 2~K. At 14~T, the magnetoresistance of S2 reaches 8590~\%. With the removal of the magnetoresistance background, quantum oscillations can be seen at high fields, as displayed in the inset of Fig.~\ref{fig4}(b). The main panel of Fig.~\ref{fig4}(b) shows the fast Fourier transform of the oscillatory data, resulting in a well-defined quantum oscillation frequency. The frequency of 261~T is consistent with an electron Fermi pocket reported earlier in free-standing MoTe$_2$~\cite{Qi2016,Rhodes2017,Hu2020,Liu2020}. Thus, the biaxial strain of this magnitude does not modify the electron pocket significantly. Furthermore, when the temperature increases, the oscillation amplitude decreases, consistent with the expected behaviour for quantum oscillations. These observations confirm that the quality of the biaxial device remains high at cryogenic temperatures, and S2 experiences a uniform strain environment.

As mentioned earlier, hydrostatic pressure has been successful in materials tuning. Then, what is the advantage of the proposed biaxial strain tuning? The most significant advantage is simplicity. The biaxial setup is significantly simpler and cheaper than a typical pressure cell. Another major advantage is the issue of accessibility -- the biaxial strain setup has an open architecture, making it compatible with many experiments. For instance, biaxial strain setup offers the possibility of conducting photoemission measurements while standard high-pressure devices do not. Next, hydrostatic pressure is always compressive; while with a careful selection of the substrate, biaxial strain can be either tensile or compressive. Finally, the substrate being insulating makes it possible to examine the biaxially strained materials using a pulsed magnet, a major boost for magnetotransport and the detection of magnetic quantum oscillations. 

In summary, we have used magnetotransport to investigate the superconductivity of MoTe$_2$ under biaxial strain with a simple device that utilizes the thermal expansion mismatch between dissimilar materials. We observe a five-fold increase in the superconducting transition temperature $T_c$, from $\sim 130$~mK to $\sim640$~mK, upon the application of a modest biaxial strain. The temperature dependence of the upper critical field $H_{c2}$ can be fairly well-described by the Werthamer-Helfand-Hohenberg model in the one-band dirty limit with null $\alpha$ and $\lambda$, indicating that the orbital effect is the dominant pair-breaking mechanism in biaxially strained MoTe$_2$. The angular dependence of $H_{c2}$ reveals a three-dimensional superconductivity in biaxially strained MoTe$_2$, despite a relatively large anisotropy factor ($\gamma=7$). The large enhancement of $T_c$ and $H_{c2}$ over the corresponding values in the free-standing MoTe$_2$ demonstrates the significant impact on the superconductivity of MoTe$_2$ by biaxial strain. The detection of quantum oscillations further confirms that the device quality is high and the strain experienced by the sample is rather uniform. Our study demonstrates that biaxial strain is a powerful tool to manipulate quasi-2D materials, and the simplicity of the setup promises applications in tuning other topical systems, such as transition metal dichalcogenides~\cite{Frindt1972,Sipos2008,Ye2012,Taniguchi2012,Kang2015,Pan2015}, iron-based superconductors~\cite{Kamihara2008,Paglione2010,Hosono2015} and recently discovered kagome superconductors AV$_3$Sb$_5$ (A = K, Rb, Cs) \cite{Ortiz2019,Ortiz2020,Neupert2021}. 

\section*{Supplementary Material}

See the supplementary material for the raw data used for the determination of $T_c$ and $H_{c2}$ for free standing MoTe$_2$ (S1) and biaxially strained MoTe$_2$ (S2), and for the estimation of biaxial strain $\varepsilon_{biaxial}$ on S2.

\begin{acknowledgments}
We acknowledge financial support from Research Grants Council of Hong Kong (GRF/14300419, GRF/14301020 and A-CUHK402/19), CUHK Direct Grant (4053463, 4053528, 4053408 and 4053461), the National Natural Science Foundation of China (12104384) 

\end{acknowledgments}

\section*{AUTHOR DECLARATIONS
}
\subsection*{{Conflict of Interest
}}
The authors have no conflicts to disclose.

\section*{Data Availability Statement}
The data that support the findings of
this study are available from the
corresponding author upon reasonable
request.


\begin{thebibliography}{41}%
\makeatletter
\providecommand \@ifxundefined [1]{%
 \@ifx{#1\undefined}
}%
\providecommand \@ifnum [1]{%
 \ifnum #1\expandafter \@firstoftwo
 \else \expandafter \@secondoftwo
 \fi
}%
\providecommand \@ifx [1]{%
 \ifx #1\expandafter \@firstoftwo
 \else \expandafter \@secondoftwo
 \fi
}%
\providecommand \natexlab [1]{#1}%
\providecommand \enquote  [1]{``#1''}%
\providecommand \bibnamefont  [1]{#1}%
\providecommand \bibfnamefont [1]{#1}%
\providecommand \citenamefont [1]{#1}%
\providecommand \href@noop [0]{\@secondoftwo}%
\providecommand \href [0]{\begingroup \@sanitize@url \@href}%
\providecommand \@href[1]{\@@startlink{#1}\@@href}%
\providecommand \@@href[1]{\endgroup#1\@@endlink}%
\providecommand \@sanitize@url [0]{\catcode `\\12\catcode `\$12\catcode
  `\&12\catcode `\#12\catcode `\^12\catcode `\_12\catcode `\%12\relax}%
\providecommand \@@startlink[1]{}%
\providecommand \@@endlink[0]{}%
\providecommand \url  [0]{\begingroup\@sanitize@url \@url }%
\providecommand \@url [1]{\endgroup\@href {#1}{\urlprefix }}%
\providecommand \urlprefix  [0]{URL }%
\providecommand \Eprint [0]{\href }%
\providecommand \doibase [0]{https://doi.org/}%
\providecommand \selectlanguage [0]{\@gobble}%
\providecommand \bibinfo  [0]{\@secondoftwo}%
\providecommand \bibfield  [0]{\@secondoftwo}%
\providecommand \translation [1]{[#1]}%
\providecommand \BibitemOpen [0]{}%
\providecommand \bibitemStop [0]{}%
\providecommand \bibitemNoStop [0]{.\EOS\space}%
\providecommand \EOS [0]{\spacefactor3000\relax}%
\providecommand \BibitemShut  [1]{\csname bibitem#1\endcsname}%
\let\auto@bib@innerbib\@empty
\bibitem [{\citenamefont {Sato}\ and\ \citenamefont {Ando}(2017)}]{Sato2017}%
  \BibitemOpen
  \bibfield  {author} {\bibinfo {author} {\bibfnamefont {M.}~\bibnamefont
  {Sato}}\ and\ \bibinfo {author} {\bibfnamefont {Y.}~\bibnamefont {Ando}},\
  }\bibfield  {title} {\enquote {\bibinfo {title} {Topological superconductors:
  a review},}\ }\href {http://stacks.iop.org/0034-4885/80/i=7/a=076501}
  {\bibfield  {journal} {\bibinfo  {journal} {Rep. Prog. Phys.}\ }\textbf
  {\bibinfo {volume} {80}},\ \bibinfo {pages} {076501} (\bibinfo {year}
  {2017})}\BibitemShut {NoStop}%
\bibitem [{\citenamefont {Li}\ and\ \citenamefont {Xu}(2019)}]{Li2019aqt}%
  \BibitemOpen
  \bibfield  {author} {\bibinfo {author} {\bibfnamefont {Y.}~\bibnamefont
  {Li}}\ and\ \bibinfo {author} {\bibfnamefont {Z.-A.}\ \bibnamefont {Xu}},\
  }\bibfield  {title} {\enquote {\bibinfo {title} {Exploring topological
  superconductivity in topological materials},}\ }\href
  {https://doi.org/https://doi.org/10.1002/qute.201800112} {\bibfield
  {journal} {\bibinfo  {journal} {Adv. Quantum Technol.}\ }\textbf {\bibinfo
  {volume} {2}},\ \bibinfo {pages} {1800112} (\bibinfo {year}
  {2019})}\BibitemShut {NoStop}%
\bibitem [{\citenamefont {Soluyanov}\ \emph {et~al.}(2015)\citenamefont
  {Soluyanov}, \citenamefont {Gresch}, \citenamefont {Wang}, \citenamefont
  {Wu}, \citenamefont {Troyer}, \citenamefont {Dai},\ and\ \citenamefont
  {Bernevig}}]{Soluyanov2015}%
  \BibitemOpen
  \bibfield  {author} {\bibinfo {author} {\bibfnamefont {A.~A.}\ \bibnamefont
  {Soluyanov}}, \bibinfo {author} {\bibfnamefont {D.}~\bibnamefont {Gresch}},
  \bibinfo {author} {\bibfnamefont {Z.}~\bibnamefont {Wang}}, \bibinfo {author}
  {\bibfnamefont {Q.}~\bibnamefont {Wu}}, \bibinfo {author} {\bibfnamefont
  {M.}~\bibnamefont {Troyer}}, \bibinfo {author} {\bibfnamefont
  {X.}~\bibnamefont {Dai}},\ and\ \bibinfo {author} {\bibfnamefont {B.~A.}\
  \bibnamefont {Bernevig}},\ }\bibfield  {title} {\enquote {\bibinfo {title}
  {Type-{II Weyl} semimetals},}\ }\href@noop {} {\bibfield  {journal} {\bibinfo
   {journal} {Nature}\ }\textbf {\bibinfo {volume} {527}},\ \bibinfo {pages}
  {495--498} (\bibinfo {year} {2015})}\BibitemShut {NoStop}%
\bibitem [{\citenamefont {Deng}\ \emph {et~al.}(2016)\citenamefont {Deng},
  \citenamefont {Wan}, \citenamefont {Deng}, \citenamefont {Zhang},
  \citenamefont {Ding}, \citenamefont {Wang}, \citenamefont {Yan},
  \citenamefont {Huang}, \citenamefont {Zhang}, \citenamefont {Xu},
  \citenamefont {Denlinger}, \citenamefont {Fedorov}, \citenamefont {Yang},
  \citenamefont {Duan}, \citenamefont {Yao}, \citenamefont {Wu}, \citenamefont
  {Fan}, \citenamefont {Zhang}, \citenamefont {Chen},\ and\ \citenamefont
  {Zhou}}]{Deng2016}%
  \BibitemOpen
  \bibfield  {author} {\bibinfo {author} {\bibfnamefont {K.}~\bibnamefont
  {Deng}}, \bibinfo {author} {\bibfnamefont {G.}~\bibnamefont {Wan}}, \bibinfo
  {author} {\bibfnamefont {P.}~\bibnamefont {Deng}}, \bibinfo {author}
  {\bibfnamefont {K.}~\bibnamefont {Zhang}}, \bibinfo {author} {\bibfnamefont
  {S.}~\bibnamefont {Ding}}, \bibinfo {author} {\bibfnamefont {E.}~\bibnamefont
  {Wang}}, \bibinfo {author} {\bibfnamefont {M.}~\bibnamefont {Yan}}, \bibinfo
  {author} {\bibfnamefont {H.}~\bibnamefont {Huang}}, \bibinfo {author}
  {\bibfnamefont {H.}~\bibnamefont {Zhang}}, \bibinfo {author} {\bibfnamefont
  {Z.}~\bibnamefont {Xu}}, \bibinfo {author} {\bibfnamefont {J.}~\bibnamefont
  {Denlinger}}, \bibinfo {author} {\bibfnamefont {A.}~\bibnamefont {Fedorov}},
  \bibinfo {author} {\bibfnamefont {H.}~\bibnamefont {Yang}}, \bibinfo {author}
  {\bibfnamefont {W.}~\bibnamefont {Duan}}, \bibinfo {author} {\bibfnamefont
  {H.}~\bibnamefont {Yao}}, \bibinfo {author} {\bibfnamefont {Y.}~\bibnamefont
  {Wu}}, \bibinfo {author} {\bibfnamefont {S.}~\bibnamefont {Fan}}, \bibinfo
  {author} {\bibfnamefont {H.}~\bibnamefont {Zhang}}, \bibinfo {author}
  {\bibfnamefont {X.}~\bibnamefont {Chen}},\ and\ \bibinfo {author}
  {\bibfnamefont {S.}~\bibnamefont {Zhou}},\ }\bibfield  {title} {\enquote
  {\bibinfo {title} {Experimental observation of topological {Fermi} arcs in
  type-{II Weyl} semimetal {MoTe$_2$}},}\ }\href@noop {} {\bibfield  {journal}
  {\bibinfo  {journal} {Nat. Phys.}\ }\textbf {\bibinfo {volume} {12}},\
  \bibinfo {pages} {1105--1110} (\bibinfo {year} {2016})}\BibitemShut {NoStop}%
\bibitem [{\citenamefont {Huang}\ \emph {et~al.}(2016)\citenamefont {Huang},
  \citenamefont {McCormick}, \citenamefont {Ochi}, \citenamefont {Zhao},
  \citenamefont {Suzuki}, \citenamefont {Arita}, \citenamefont {Wu},
  \citenamefont {Mou}, \citenamefont {Cao}, \citenamefont {Yan}, \citenamefont
  {Trivedi},\ and\ \citenamefont {Kaminski}}]{Huang2016}%
  \BibitemOpen
  \bibfield  {author} {\bibinfo {author} {\bibfnamefont {L.}~\bibnamefont
  {Huang}}, \bibinfo {author} {\bibfnamefont {T.~M.}\ \bibnamefont
  {McCormick}}, \bibinfo {author} {\bibfnamefont {M.}~\bibnamefont {Ochi}},
  \bibinfo {author} {\bibfnamefont {Z.}~\bibnamefont {Zhao}}, \bibinfo {author}
  {\bibfnamefont {M.-T.}\ \bibnamefont {Suzuki}}, \bibinfo {author}
  {\bibfnamefont {R.}~\bibnamefont {Arita}}, \bibinfo {author} {\bibfnamefont
  {Y.}~\bibnamefont {Wu}}, \bibinfo {author} {\bibfnamefont {D.}~\bibnamefont
  {Mou}}, \bibinfo {author} {\bibfnamefont {H.}~\bibnamefont {Cao}}, \bibinfo
  {author} {\bibfnamefont {J.}~\bibnamefont {Yan}}, \bibinfo {author}
  {\bibfnamefont {N.}~\bibnamefont {Trivedi}},\ and\ \bibinfo {author}
  {\bibfnamefont {A.}~\bibnamefont {Kaminski}},\ }\bibfield  {title} {\enquote
  {\bibinfo {title} {Spectroscopic evidence for a type {II Weyl} semimetallic
  state in {MoTe$_2$}},}\ }\href {https://doi.org/10.1038/nmat4685} {\bibfield
  {journal} {\bibinfo  {journal} {Nat. Mater.}\ }\textbf {\bibinfo {volume}
  {15}},\ \bibinfo {pages} {1155--1160} (\bibinfo {year} {2016})}\BibitemShut
  {NoStop}%
\bibitem [{\citenamefont {Jiang}\ \emph {et~al.}(2017)\citenamefont {Jiang},
  \citenamefont {Liu}, \citenamefont {Sun}, \citenamefont {Yang}, \citenamefont
  {Rajamathi}, \citenamefont {Qi}, \citenamefont {Yang}, \citenamefont {Chen},
  \citenamefont {Peng}, \citenamefont {Hwang}, \citenamefont {Sun},
  \citenamefont {Mo}, \citenamefont {Vobornik}, \citenamefont {Fujii},
  \citenamefont {Parkin}, \citenamefont {Felser}, \citenamefont {Yan},\ and\
  \citenamefont {Chen}}]{Jiang2017nc}%
  \BibitemOpen
  \bibfield  {author} {\bibinfo {author} {\bibfnamefont {J.}~\bibnamefont
  {Jiang}}, \bibinfo {author} {\bibfnamefont {Z.~K.}\ \bibnamefont {Liu}},
  \bibinfo {author} {\bibfnamefont {Y.}~\bibnamefont {Sun}}, \bibinfo {author}
  {\bibfnamefont {H.~F.}\ \bibnamefont {Yang}}, \bibinfo {author}
  {\bibfnamefont {C.~R.}\ \bibnamefont {Rajamathi}}, \bibinfo {author}
  {\bibfnamefont {Y.~P.}\ \bibnamefont {Qi}}, \bibinfo {author} {\bibfnamefont
  {L.~X.}\ \bibnamefont {Yang}}, \bibinfo {author} {\bibfnamefont
  {C.}~\bibnamefont {Chen}}, \bibinfo {author} {\bibfnamefont {H.}~\bibnamefont
  {Peng}}, \bibinfo {author} {\bibfnamefont {C.-C.}\ \bibnamefont {Hwang}},
  \bibinfo {author} {\bibfnamefont {S.~Z.}\ \bibnamefont {Sun}}, \bibinfo
  {author} {\bibfnamefont {S.-K.}\ \bibnamefont {Mo}}, \bibinfo {author}
  {\bibfnamefont {I.}~\bibnamefont {Vobornik}}, \bibinfo {author}
  {\bibfnamefont {J.}~\bibnamefont {Fujii}}, \bibinfo {author} {\bibfnamefont
  {S.~S.~P.}\ \bibnamefont {Parkin}}, \bibinfo {author} {\bibfnamefont
  {C.}~\bibnamefont {Felser}}, \bibinfo {author} {\bibfnamefont {B.~H.}\
  \bibnamefont {Yan}},\ and\ \bibinfo {author} {\bibfnamefont {Y.~L.}\
  \bibnamefont {Chen}},\ }\bibfield  {title} {\enquote {\bibinfo {title}
  {{Signature of type-II Weyl semimetal phase in MoTe$_2$}},}\ }\href@noop {}
  {\bibfield  {journal} {\bibinfo  {journal} {Nat. Commun.}\ }\textbf {\bibinfo
  {volume} {8}},\ \bibinfo {pages} {13973} (\bibinfo {year}
  {2017})}\BibitemShut {NoStop}%
\bibitem [{\citenamefont {Wang}\ \emph {et~al.}(2020)\citenamefont {Wang},
  \citenamefont {Kim}, \citenamefont {Liu}, \citenamefont {Cevallos},
  \citenamefont {Cava},\ and\ \citenamefont {Ong}}]{Wang2020}%
  \BibitemOpen
  \bibfield  {author} {\bibinfo {author} {\bibfnamefont {W.}~\bibnamefont
  {Wang}}, \bibinfo {author} {\bibfnamefont {S.}~\bibnamefont {Kim}}, \bibinfo
  {author} {\bibfnamefont {M.}~\bibnamefont {Liu}}, \bibinfo {author}
  {\bibfnamefont {F.~A.}\ \bibnamefont {Cevallos}}, \bibinfo {author}
  {\bibfnamefont {R.~J.}\ \bibnamefont {Cava}},\ and\ \bibinfo {author}
  {\bibfnamefont {N.~P.}\ \bibnamefont {Ong}},\ }\bibfield  {title} {\enquote
  {\bibinfo {title} {Evidence for an edge supercurrent in the {Weyl}
  superconductor {MoTe$_2$}},}\ }\href
  {https://doi.org/10.1126/science.aaw9270} {\bibfield  {journal} {\bibinfo
  {journal} {Science}\ }\textbf {\bibinfo {volume} {368}},\ \bibinfo {pages}
  {534--537} (\bibinfo {year} {2020})}\BibitemShut {NoStop}%
\bibitem [{\citenamefont {Qi}\ \emph {et~al.}(2016)\citenamefont {Qi},
  \citenamefont {Naumov}, \citenamefont {Ali}, \citenamefont {Rajamathi},
  \citenamefont {Schnelle}, \citenamefont {Barkalov}, \citenamefont {Hanfland},
  \citenamefont {Wu}, \citenamefont {Shekhar}, \citenamefont {Sun},
  \citenamefont {S{\"u}{\ss}}, \citenamefont {Schmidt}, \citenamefont
  {Schwarz}, \citenamefont {Pippel}, \citenamefont {Werner}, \citenamefont
  {Hillebrand}, \citenamefont {F{\"o}rster}, \citenamefont {Kampert},
  \citenamefont {Parkin}, \citenamefont {Cava}, \citenamefont {Felser},
  \citenamefont {Yan},\ and\ \citenamefont {Medvedev}}]{Qi2016}%
  \BibitemOpen
  \bibfield  {author} {\bibinfo {author} {\bibfnamefont {Y.}~\bibnamefont
  {Qi}}, \bibinfo {author} {\bibfnamefont {P.~G.}\ \bibnamefont {Naumov}},
  \bibinfo {author} {\bibfnamefont {M.~N.}\ \bibnamefont {Ali}}, \bibinfo
  {author} {\bibfnamefont {C.~R.}\ \bibnamefont {Rajamathi}}, \bibinfo {author}
  {\bibfnamefont {W.}~\bibnamefont {Schnelle}}, \bibinfo {author}
  {\bibfnamefont {O.}~\bibnamefont {Barkalov}}, \bibinfo {author}
  {\bibfnamefont {M.}~\bibnamefont {Hanfland}}, \bibinfo {author}
  {\bibfnamefont {S.-C.}\ \bibnamefont {Wu}}, \bibinfo {author} {\bibfnamefont
  {C.}~\bibnamefont {Shekhar}}, \bibinfo {author} {\bibfnamefont
  {Y.}~\bibnamefont {Sun}}, \bibinfo {author} {\bibfnamefont {V.}~\bibnamefont
  {S{\"u}{\ss}}}, \bibinfo {author} {\bibfnamefont {M.}~\bibnamefont
  {Schmidt}}, \bibinfo {author} {\bibfnamefont {U.}~\bibnamefont {Schwarz}},
  \bibinfo {author} {\bibfnamefont {E.}~\bibnamefont {Pippel}}, \bibinfo
  {author} {\bibfnamefont {P.}~\bibnamefont {Werner}}, \bibinfo {author}
  {\bibfnamefont {R.}~\bibnamefont {Hillebrand}}, \bibinfo {author}
  {\bibfnamefont {T.}~\bibnamefont {F{\"o}rster}}, \bibinfo {author}
  {\bibfnamefont {E.}~\bibnamefont {Kampert}}, \bibinfo {author} {\bibfnamefont
  {S.}~\bibnamefont {Parkin}}, \bibinfo {author} {\bibfnamefont {R.~J.}\
  \bibnamefont {Cava}}, \bibinfo {author} {\bibfnamefont {C.}~\bibnamefont
  {Felser}}, \bibinfo {author} {\bibfnamefont {B.}~\bibnamefont {Yan}},\ and\
  \bibinfo {author} {\bibfnamefont {S.~A.}\ \bibnamefont {Medvedev}},\
  }\bibfield  {title} {\enquote {\bibinfo {title} {Superconductivity in {Weyl}
  semimetal candidate {MoTe$_2$}},}\ }\href
  {http://dx.doi.org/10.1038/ncomms11038} {\bibfield  {journal} {\bibinfo
  {journal} {Nat. Commun.}\ }\textbf {\bibinfo {volume} {7}},\ \bibinfo {pages}
  {11038} (\bibinfo {year} {2016})}\BibitemShut {NoStop}%
\bibitem [{\citenamefont {Hu}\ \emph {et~al.}(2019)\citenamefont {Hu},
  \citenamefont {Chan}, \citenamefont {Lai}, \citenamefont {Ho}, \citenamefont
  {Guo}, \citenamefont {Sun}, \citenamefont {Yip}, \citenamefont {Ng},
  \citenamefont {Lu},\ and\ \citenamefont {Goh}}]{Hu2019}%
  \BibitemOpen
  \bibfield  {author} {\bibinfo {author} {\bibfnamefont {Y.~J.}\ \bibnamefont
  {Hu}}, \bibinfo {author} {\bibfnamefont {Y.~T.}\ \bibnamefont {Chan}},
  \bibinfo {author} {\bibfnamefont {K.~T.}\ \bibnamefont {Lai}}, \bibinfo
  {author} {\bibfnamefont {K.~O.}\ \bibnamefont {Ho}}, \bibinfo {author}
  {\bibfnamefont {X.}~\bibnamefont {Guo}}, \bibinfo {author} {\bibfnamefont
  {H.-P.}\ \bibnamefont {Sun}}, \bibinfo {author} {\bibfnamefont {K.~Y.}\
  \bibnamefont {Yip}}, \bibinfo {author} {\bibfnamefont {D.~H.~L.}\
  \bibnamefont {Ng}}, \bibinfo {author} {\bibfnamefont {H.-Z.}\ \bibnamefont
  {Lu}},\ and\ \bibinfo {author} {\bibfnamefont {S.~K.}\ \bibnamefont {Goh}},\
  }\bibfield  {title} {\enquote {\bibinfo {title} {Angular dependence of the
  upper critical field in the high-pressure {$1{T}^{\ensuremath{'}}$} phase of
  {${\mathrm{MoTe}}_{2}$}},}\ }\href
  {https://doi.org/10.1103/PhysRevMaterials.3.034201} {\bibfield  {journal}
  {\bibinfo  {journal} {Phys. Rev. Mater.}\ }\textbf {\bibinfo {volume} {3}},\
  \bibinfo {pages} {034201} (\bibinfo {year} {2019})}\BibitemShut {NoStop}%
\bibitem [{\citenamefont {Luo}\ \emph {et~al.}(2020)\citenamefont {Luo},
  \citenamefont {Li}, \citenamefont {Zhang}, \citenamefont {Ji}, \citenamefont
  {Wang}, \citenamefont {Shan}, \citenamefont {Zhang}, \citenamefont {Cai},
  \citenamefont {Liu}, \citenamefont {Wang}, \citenamefont {Zhang},\ and\
  \citenamefont {Wang}}]{Luo2020}%
  \BibitemOpen
  \bibfield  {author} {\bibinfo {author} {\bibfnamefont {J.}~\bibnamefont
  {Luo}}, \bibinfo {author} {\bibfnamefont {Y.}~\bibnamefont {Li}}, \bibinfo
  {author} {\bibfnamefont {J.}~\bibnamefont {Zhang}}, \bibinfo {author}
  {\bibfnamefont {H.}~\bibnamefont {Ji}}, \bibinfo {author} {\bibfnamefont
  {H.}~\bibnamefont {Wang}}, \bibinfo {author} {\bibfnamefont {J.-Y.}\
  \bibnamefont {Shan}}, \bibinfo {author} {\bibfnamefont {C.}~\bibnamefont
  {Zhang}}, \bibinfo {author} {\bibfnamefont {C.}~\bibnamefont {Cai}}, \bibinfo
  {author} {\bibfnamefont {J.}~\bibnamefont {Liu}}, \bibinfo {author}
  {\bibfnamefont {Y.}~\bibnamefont {Wang}}, \bibinfo {author} {\bibfnamefont
  {Y.}~\bibnamefont {Zhang}},\ and\ \bibinfo {author} {\bibfnamefont
  {J.}~\bibnamefont {Wang}},\ }\bibfield  {title} {\enquote {\bibinfo {title}
  {Possible unconventional two-band superconductivity in
  {$\mathrm{Mo}{\mathrm{Te}}_{2}$}},}\ }\href
  {https://doi.org/10.1103/PhysRevB.102.064502} {\bibfield  {journal} {\bibinfo
   {journal} {Phys. Rev. B}\ }\textbf {\bibinfo {volume} {102}},\ \bibinfo
  {pages} {064502} (\bibinfo {year} {2020})}\BibitemShut {NoStop}%
\bibitem [{\citenamefont {Takahashi}\ \emph {et~al.}(2017)\citenamefont
  {Takahashi}, \citenamefont {Akiba}, \citenamefont {Imura}, \citenamefont
  {Shiino}, \citenamefont {Deguchi}, \citenamefont {Sato}, \citenamefont
  {Sakai}, \citenamefont {Bahramy},\ and\ \citenamefont
  {Ishiwata}}]{Takahashi2017}%
  \BibitemOpen
  \bibfield  {author} {\bibinfo {author} {\bibfnamefont {H.}~\bibnamefont
  {Takahashi}}, \bibinfo {author} {\bibfnamefont {T.}~\bibnamefont {Akiba}},
  \bibinfo {author} {\bibfnamefont {K.}~\bibnamefont {Imura}}, \bibinfo
  {author} {\bibfnamefont {T.}~\bibnamefont {Shiino}}, \bibinfo {author}
  {\bibfnamefont {K.}~\bibnamefont {Deguchi}}, \bibinfo {author} {\bibfnamefont
  {N.~K.}\ \bibnamefont {Sato}}, \bibinfo {author} {\bibfnamefont
  {H.}~\bibnamefont {Sakai}}, \bibinfo {author} {\bibfnamefont {M.~S.}\
  \bibnamefont {Bahramy}},\ and\ \bibinfo {author} {\bibfnamefont
  {S.}~\bibnamefont {Ishiwata}},\ }\bibfield  {title} {\enquote {\bibinfo
  {title} {Anticorrelation between polar lattice instability and
  superconductivity in the {Weyl} semimetal candidate
  {${\mathrm{MoTe}}_{2}$}},}\ }\href
  {https://doi.org/10.1103/PhysRevB.95.100501} {\bibfield  {journal} {\bibinfo
  {journal} {Phys. Rev. B}\ }\textbf {\bibinfo {volume} {95}},\ \bibinfo
  {pages} {100501} (\bibinfo {year} {2017})}\BibitemShut {NoStop}%
\bibitem [{\citenamefont {Heikes}\ \emph {et~al.}(2018)\citenamefont {Heikes},
  \citenamefont {Liu}, \citenamefont {Metz}, \citenamefont {Eckberg},
  \citenamefont {Neves}, \citenamefont {Wu}, \citenamefont {Hung},
  \citenamefont {Piccoli}, \citenamefont {Cao}, \citenamefont {Leao},
  \citenamefont {Paglione}, \citenamefont {Yildirim}, \citenamefont {Butch},\
  and\ \citenamefont {Ratcliff}}]{Heikes2018}%
  \BibitemOpen
  \bibfield  {author} {\bibinfo {author} {\bibfnamefont {C.}~\bibnamefont
  {Heikes}}, \bibinfo {author} {\bibfnamefont {I.-L.}\ \bibnamefont {Liu}},
  \bibinfo {author} {\bibfnamefont {T.}~\bibnamefont {Metz}}, \bibinfo {author}
  {\bibfnamefont {C.}~\bibnamefont {Eckberg}}, \bibinfo {author} {\bibfnamefont
  {P.}~\bibnamefont {Neves}}, \bibinfo {author} {\bibfnamefont
  {Y.}~\bibnamefont {Wu}}, \bibinfo {author} {\bibfnamefont {L.}~\bibnamefont
  {Hung}}, \bibinfo {author} {\bibfnamefont {P.}~\bibnamefont {Piccoli}},
  \bibinfo {author} {\bibfnamefont {H.}~\bibnamefont {Cao}}, \bibinfo {author}
  {\bibfnamefont {J.}~\bibnamefont {Leao}}, \bibinfo {author} {\bibfnamefont
  {J.}~\bibnamefont {Paglione}}, \bibinfo {author} {\bibfnamefont
  {T.}~\bibnamefont {Yildirim}}, \bibinfo {author} {\bibfnamefont {N.~P.}\
  \bibnamefont {Butch}},\ and\ \bibinfo {author} {\bibfnamefont
  {W.}~\bibnamefont {Ratcliff}},\ }\bibfield  {title} {\enquote {\bibinfo
  {title} {Mechanical control of crystal symmetry and superconductivity in
  {Weyl} semimetal {${\mathrm{MoTe}}_{2}$}},}\ }\href
  {https://doi.org/10.1103/PhysRevMaterials.2.074202} {\bibfield  {journal}
  {\bibinfo  {journal} {Phys. Rev. Mater.}\ }\textbf {\bibinfo {volume} {2}},\
  \bibinfo {pages} {074202} (\bibinfo {year} {2018})}\BibitemShut {NoStop}%
\bibitem [{\citenamefont {Lee}\ \emph {et~al.}(2018)\citenamefont {Lee},
  \citenamefont {Jang}, \citenamefont {Kim}, \citenamefont {Jung},
  \citenamefont {Kim}, \citenamefont {Cho}, \citenamefont {Kim}, \citenamefont
  {Rhee}, \citenamefont {Park},\ and\ \citenamefont {Park}}]{Lee2018}%
  \BibitemOpen
  \bibfield  {author} {\bibinfo {author} {\bibfnamefont {S.}~\bibnamefont
  {Lee}}, \bibinfo {author} {\bibfnamefont {J.}~\bibnamefont {Jang}}, \bibinfo
  {author} {\bibfnamefont {S.-I.}\ \bibnamefont {Kim}}, \bibinfo {author}
  {\bibfnamefont {S.-G.}\ \bibnamefont {Jung}}, \bibinfo {author}
  {\bibfnamefont {J.}~\bibnamefont {Kim}}, \bibinfo {author} {\bibfnamefont
  {S.}~\bibnamefont {Cho}}, \bibinfo {author} {\bibfnamefont {S.~W.}\
  \bibnamefont {Kim}}, \bibinfo {author} {\bibfnamefont {J.~Y.}\ \bibnamefont
  {Rhee}}, \bibinfo {author} {\bibfnamefont {K.-S.}\ \bibnamefont {Park}},\
  and\ \bibinfo {author} {\bibfnamefont {T.}~\bibnamefont {Park}},\ }\bibfield
  {title} {\enquote {\bibinfo {title} {Origin of extremely large
  magnetoresistance in the candidate type-{II Weyl} semimetal
  {MoTe$_{2-x}$}},}\ }\href {https://doi.org/10.1038/s41598-018-32387-1}
  {\bibfield  {journal} {\bibinfo  {journal} {Sci. Rep.}\ }\textbf {\bibinfo
  {volume} {8}},\ \bibinfo {pages} {13937} (\bibinfo {year}
  {2018})}\BibitemShut {NoStop}%
\bibitem [{\citenamefont {Guguchia}\ \emph {et~al.}(2017)\citenamefont
  {Guguchia}, \citenamefont {von Rohr}, \citenamefont {Shermadini},
  \citenamefont {Lee}, \citenamefont {Banerjee}, \citenamefont {Wieteska},
  \citenamefont {Marianetti}, \citenamefont {Frandsen}, \citenamefont
  {Luetkens}, \citenamefont {Gong}, \citenamefont {Cheung}, \citenamefont
  {Baines}, \citenamefont {Shengelaya}, \citenamefont {Taniashvili},
  \citenamefont {Pasupathy}, \citenamefont {Morenzoni}, \citenamefont
  {Billinge}, \citenamefont {Amato}, \citenamefont {Cava}, \citenamefont
  {Khasanov},\ and\ \citenamefont {Uemura}}]{Guguchia2017}%
  \BibitemOpen
  \bibfield  {author} {\bibinfo {author} {\bibfnamefont {Z.}~\bibnamefont
  {Guguchia}}, \bibinfo {author} {\bibfnamefont {F.}~\bibnamefont {von Rohr}},
  \bibinfo {author} {\bibfnamefont {Z.}~\bibnamefont {Shermadini}}, \bibinfo
  {author} {\bibfnamefont {A.~T.}\ \bibnamefont {Lee}}, \bibinfo {author}
  {\bibfnamefont {S.}~\bibnamefont {Banerjee}}, \bibinfo {author}
  {\bibfnamefont {A.~R.}\ \bibnamefont {Wieteska}}, \bibinfo {author}
  {\bibfnamefont {C.~A.}\ \bibnamefont {Marianetti}}, \bibinfo {author}
  {\bibfnamefont {B.~A.}\ \bibnamefont {Frandsen}}, \bibinfo {author}
  {\bibfnamefont {H.}~\bibnamefont {Luetkens}}, \bibinfo {author}
  {\bibfnamefont {Z.}~\bibnamefont {Gong}}, \bibinfo {author} {\bibfnamefont
  {S.~C.}\ \bibnamefont {Cheung}}, \bibinfo {author} {\bibfnamefont
  {C.}~\bibnamefont {Baines}}, \bibinfo {author} {\bibfnamefont
  {A.}~\bibnamefont {Shengelaya}}, \bibinfo {author} {\bibfnamefont
  {G.}~\bibnamefont {Taniashvili}}, \bibinfo {author} {\bibfnamefont {A.~N.}\
  \bibnamefont {Pasupathy}}, \bibinfo {author} {\bibfnamefont {E.}~\bibnamefont
  {Morenzoni}}, \bibinfo {author} {\bibfnamefont {S.~J.~L.}\ \bibnamefont
  {Billinge}}, \bibinfo {author} {\bibfnamefont {A.}~\bibnamefont {Amato}},
  \bibinfo {author} {\bibfnamefont {R.~J.}\ \bibnamefont {Cava}}, \bibinfo
  {author} {\bibfnamefont {R.}~\bibnamefont {Khasanov}},\ and\ \bibinfo
  {author} {\bibfnamefont {Y.~J.}\ \bibnamefont {Uemura}},\ }\bibfield  {title}
  {\enquote {\bibinfo {title} {Signatures of the topological s$^{+-}$
  superconducting order parameter in the type-{II Weyl} semimetal
  {$T_d$-MoTe$_2$}},}\ }\href {https://doi.org/10.1038/s41467-017-01066-6}
  {\bibfield  {journal} {\bibinfo  {journal} {Nat. Commun.}\ }\textbf {\bibinfo
  {volume} {8}},\ \bibinfo {pages} {1082} (\bibinfo {year} {2017})}\BibitemShut
  {NoStop}%
\bibitem [{\citenamefont {Chen}\ \emph {et~al.}(2016)\citenamefont {Chen},
  \citenamefont {Luo}, \citenamefont {Xiao}, \citenamefont {Lu}, \citenamefont
  {Zhang}, \citenamefont {Yang}, \citenamefont {Li}, \citenamefont {Pei},
  \citenamefont {Shao}, \citenamefont {Zhang}, \citenamefont {Ling},
  \citenamefont {Xi}, \citenamefont {Song},\ and\ \citenamefont
  {Sun}}]{Chen2016}%
  \BibitemOpen
  \bibfield  {author} {\bibinfo {author} {\bibfnamefont {F.~C.}\ \bibnamefont
  {Chen}}, \bibinfo {author} {\bibfnamefont {X.}~\bibnamefont {Luo}}, \bibinfo
  {author} {\bibfnamefont {R.~C.}\ \bibnamefont {Xiao}}, \bibinfo {author}
  {\bibfnamefont {W.~J.}\ \bibnamefont {Lu}}, \bibinfo {author} {\bibfnamefont
  {B.}~\bibnamefont {Zhang}}, \bibinfo {author} {\bibfnamefont {H.~X.}\
  \bibnamefont {Yang}}, \bibinfo {author} {\bibfnamefont {J.~Q.}\ \bibnamefont
  {Li}}, \bibinfo {author} {\bibfnamefont {Q.~L.}\ \bibnamefont {Pei}},
  \bibinfo {author} {\bibfnamefont {D.~F.}\ \bibnamefont {Shao}}, \bibinfo
  {author} {\bibfnamefont {R.~R.}\ \bibnamefont {Zhang}}, \bibinfo {author}
  {\bibfnamefont {L.~S.}\ \bibnamefont {Ling}}, \bibinfo {author}
  {\bibfnamefont {C.~Y.}\ \bibnamefont {Xi}}, \bibinfo {author} {\bibfnamefont
  {W.~H.}\ \bibnamefont {Song}},\ and\ \bibinfo {author} {\bibfnamefont
  {Y.~P.}\ \bibnamefont {Sun}},\ }\bibfield  {title} {\enquote {\bibinfo
  {title} {Superconductivity enhancement in the {S-doped Weyl} semimetal
  candidate {MoTe$_2$}},}\ }\href {https://doi.org/10.1063/1.4947433}
  {\bibfield  {journal} {\bibinfo  {journal} {Appl. Phys. Lett.}\ }\textbf
  {\bibinfo {volume} {108}},\ \bibinfo {pages} {162601} (\bibinfo {year}
  {2016})}\BibitemShut {NoStop}%
\bibitem [{\citenamefont {Cho}\ \emph {et~al.}(2017)\citenamefont {Cho},
  \citenamefont {Kang}, \citenamefont {Yu}, \citenamefont {Kim}, \citenamefont
  {Ko}, \citenamefont {Hwang}, \citenamefont {Han}, \citenamefont {Choe},
  \citenamefont {Jung}, \citenamefont {Chang}, \citenamefont {Lee},
  \citenamefont {Yang},\ and\ \citenamefont {Kim}}]{Cho2017}%
  \BibitemOpen
  \bibfield  {author} {\bibinfo {author} {\bibfnamefont {S.}~\bibnamefont
  {Cho}}, \bibinfo {author} {\bibfnamefont {S.~H.}\ \bibnamefont {Kang}},
  \bibinfo {author} {\bibfnamefont {H.~S.}\ \bibnamefont {Yu}}, \bibinfo
  {author} {\bibfnamefont {H.~W.}\ \bibnamefont {Kim}}, \bibinfo {author}
  {\bibfnamefont {W.}~\bibnamefont {Ko}}, \bibinfo {author} {\bibfnamefont
  {S.~W.}\ \bibnamefont {Hwang}}, \bibinfo {author} {\bibfnamefont {W.~H.}\
  \bibnamefont {Han}}, \bibinfo {author} {\bibfnamefont {D.-H.}\ \bibnamefont
  {Choe}}, \bibinfo {author} {\bibfnamefont {Y.~H.}\ \bibnamefont {Jung}},
  \bibinfo {author} {\bibfnamefont {K.~J.}\ \bibnamefont {Chang}}, \bibinfo
  {author} {\bibfnamefont {Y.~H.}\ \bibnamefont {Lee}}, \bibinfo {author}
  {\bibfnamefont {H.}~\bibnamefont {Yang}},\ and\ \bibinfo {author}
  {\bibfnamefont {S.~W.}\ \bibnamefont {Kim}},\ }\bibfield  {title} {\enquote
  {\bibinfo {title} {Te vacancy-driven superconductivity in orthorhombic
  molybdenum ditelluride},}\ }\href
  {http://stacks.iop.org/2053-1583/4/i=2/a=021030} {\bibfield  {journal}
  {\bibinfo  {journal} {2D Mater.}\ }\textbf {\bibinfo {volume} {4}},\ \bibinfo
  {pages} {021030} (\bibinfo {year} {2017})}\BibitemShut {NoStop}%
\bibitem [{\citenamefont {Mandal}\ \emph {et~al.}(2018)\citenamefont {Mandal},
  \citenamefont {Marik}, \citenamefont {Sajilesh}, \citenamefont {Arushi},
  \citenamefont {Singh}, \citenamefont {Chakraborty}, \citenamefont {Ganguli},\
  and\ \citenamefont {Singh}}]{Mandal2018}%
  \BibitemOpen
  \bibfield  {author} {\bibinfo {author} {\bibfnamefont {M.}~\bibnamefont
  {Mandal}}, \bibinfo {author} {\bibfnamefont {S.}~\bibnamefont {Marik}},
  \bibinfo {author} {\bibfnamefont {K.~P.}\ \bibnamefont {Sajilesh}}, \bibinfo
  {author} {\bibnamefont {Arushi}}, \bibinfo {author} {\bibfnamefont
  {D.}~\bibnamefont {Singh}}, \bibinfo {author} {\bibfnamefont
  {J.}~\bibnamefont {Chakraborty}}, \bibinfo {author} {\bibfnamefont
  {N.}~\bibnamefont {Ganguli}},\ and\ \bibinfo {author} {\bibfnamefont {R.~P.}\
  \bibnamefont {Singh}},\ }\bibfield  {title} {\enquote {\bibinfo {title}
  {Enhancement of the superconducting transition temperature by {Re} doping in
  {Weyl} semimetal {${\mathrm{MoTe}}_{2}$}},}\ }\href
  {https://doi.org/10.1103/PhysRevMaterials.2.094201} {\bibfield  {journal}
  {\bibinfo  {journal} {Phys. Rev. Mater.}\ }\textbf {\bibinfo {volume} {2}},\
  \bibinfo {pages} {094201} (\bibinfo {year} {2018})}\BibitemShut {NoStop}%
\bibitem [{\citenamefont {Dahal}\ \emph {et~al.}(2020)\citenamefont {Dahal},
  \citenamefont {Deng}, \citenamefont {Poudel}, \citenamefont {Gooch},
  \citenamefont {Wu}, \citenamefont {Wu}, \citenamefont {Yang}, \citenamefont
  {Chang},\ and\ \citenamefont {Chu}}]{Dahal2020}%
  \BibitemOpen
  \bibfield  {author} {\bibinfo {author} {\bibfnamefont {R.}~\bibnamefont
  {Dahal}}, \bibinfo {author} {\bibfnamefont {L.~Z.}\ \bibnamefont {Deng}},
  \bibinfo {author} {\bibfnamefont {N.}~\bibnamefont {Poudel}}, \bibinfo
  {author} {\bibfnamefont {M.}~\bibnamefont {Gooch}}, \bibinfo {author}
  {\bibfnamefont {Z.}~\bibnamefont {Wu}}, \bibinfo {author} {\bibfnamefont
  {H.~C.}\ \bibnamefont {Wu}}, \bibinfo {author} {\bibfnamefont {H.~D.}\
  \bibnamefont {Yang}}, \bibinfo {author} {\bibfnamefont {C.~K.}\ \bibnamefont
  {Chang}},\ and\ \bibinfo {author} {\bibfnamefont {C.~W.}\ \bibnamefont
  {Chu}},\ }\bibfield  {title} {\enquote {\bibinfo {title} {Tunable structural
  phase transition and superconductivity in the {Weyl} semimetal
  {${\mathrm{Mo}}_{1\ensuremath{-}x}{\mathrm{W}}_{x}{\mathrm{Te}}_{2}$}},}\
  }\href {https://doi.org/10.1103/PhysRevB.101.140505} {\bibfield  {journal}
  {\bibinfo  {journal} {Phys. Rev. B}\ }\textbf {\bibinfo {volume} {101}},\
  \bibinfo {pages} {140505} (\bibinfo {year} {2020})}\BibitemShut {NoStop}%
\bibitem [{\citenamefont {Mandal}\ \emph {et~al.}(2021)\citenamefont {Mandal},
  \citenamefont {Patra}, \citenamefont {Kataria}, \citenamefont {Paul},
  \citenamefont {Saha},\ and\ \citenamefont {Singh}}]{Mandal2021}%
  \BibitemOpen
  \bibfield  {author} {\bibinfo {author} {\bibfnamefont {M.}~\bibnamefont
  {Mandal}}, \bibinfo {author} {\bibfnamefont {C.}~\bibnamefont {Patra}},
  \bibinfo {author} {\bibfnamefont {A.}~\bibnamefont {Kataria}}, \bibinfo
  {author} {\bibfnamefont {S.}~\bibnamefont {Paul}}, \bibinfo {author}
  {\bibfnamefont {S.}~\bibnamefont {Saha}},\ and\ \bibinfo {author}
  {\bibfnamefont {R.~P.}\ \bibnamefont {Singh}},\ }\bibfield  {title} {\enquote
  {\bibinfo {title} {Superconductivity in doped {Weyl} semimetal
  {Mo$_{0.9}$Ir$_{0.1}$Te$_2$} with broken inversion symmetry},}\ }\href
  {https://doi.org/10.1088/1361-6668/ac3b38} {\bibfield  {journal} {\bibinfo
  {journal} {Supercond. Sci. Technol.}\ }\textbf {\bibinfo {volume} {35}},\
  \bibinfo {pages} {025011} (\bibinfo {year} {2021})}\BibitemShut {NoStop}%
\bibitem [{\citenamefont {B\"ohmer}\ \emph {et~al.}(2017)\citenamefont
  {B\"ohmer}, \citenamefont {Sapkota}, \citenamefont {Kreyssig}, \citenamefont
  {Bud'ko}, \citenamefont {Drachuck}, \citenamefont {Saunders}, \citenamefont
  {Goldman},\ and\ \citenamefont {Canfield}}]{Bohmer2017}%
  \BibitemOpen
  \bibfield  {author} {\bibinfo {author} {\bibfnamefont {A.~E.}\ \bibnamefont
  {B\"ohmer}}, \bibinfo {author} {\bibfnamefont {A.}~\bibnamefont {Sapkota}},
  \bibinfo {author} {\bibfnamefont {A.}~\bibnamefont {Kreyssig}}, \bibinfo
  {author} {\bibfnamefont {S.~L.}\ \bibnamefont {Bud'ko}}, \bibinfo {author}
  {\bibfnamefont {G.}~\bibnamefont {Drachuck}}, \bibinfo {author}
  {\bibfnamefont {S.~M.}\ \bibnamefont {Saunders}}, \bibinfo {author}
  {\bibfnamefont {A.~I.}\ \bibnamefont {Goldman}},\ and\ \bibinfo {author}
  {\bibfnamefont {P.~C.}\ \bibnamefont {Canfield}},\ }\bibfield  {title}
  {\enquote {\bibinfo {title} {Effect of biaxial strain on the phase
  transitions of
  {$\mathrm{Ca}({\mathrm{Fe}}_{1\ensuremath{-}x}{\mathrm{Co}}_{x}{)}_{2}{\mathrm{As}}_{2}$}},}\
  }\href {https://doi.org/10.1103/PhysRevLett.118.107002} {\bibfield  {journal}
  {\bibinfo  {journal} {Phys. Rev. Lett.}\ }\textbf {\bibinfo {volume} {118}},\
  \bibinfo {pages} {107002} (\bibinfo {year} {2017})}\BibitemShut {NoStop}%
\bibitem [{\citenamefont {Nakajima}, \citenamefont {Ohata},\ and\ \citenamefont
  {Tajima}(2021)}]{Nakajima2021}%
  \BibitemOpen
  \bibfield  {author} {\bibinfo {author} {\bibfnamefont {M.}~\bibnamefont
  {Nakajima}}, \bibinfo {author} {\bibfnamefont {Y.}~\bibnamefont {Ohata}},\
  and\ \bibinfo {author} {\bibfnamefont {S.}~\bibnamefont {Tajima}},\
  }\bibfield  {title} {\enquote {\bibinfo {title} {Control of band structure of
  {FeSe} single crystals via biaxial strain},}\ }\href
  {https://doi.org/10.1103/PhysRevMaterials.5.044801} {\bibfield  {journal}
  {\bibinfo  {journal} {Phys. Rev. Mater.}\ }\textbf {\bibinfo {volume} {5}},\
  \bibinfo {pages} {044801} (\bibinfo {year} {2021})}\BibitemShut {NoStop}%
\bibitem [{\citenamefont {Hu}\ \emph {et~al.}(2020)\citenamefont {Hu},
  \citenamefont {Yu}, \citenamefont {Lai}, \citenamefont {Sun}, \citenamefont
  {Balakirev}, \citenamefont {Zhang}, \citenamefont {Xie}, \citenamefont {Yip},
  \citenamefont {Aulestia}, \citenamefont {Jha}, \citenamefont {Higashinaka},
  \citenamefont {Matsuda}, \citenamefont {Yanase}, \citenamefont {Aoki},\ and\
  \citenamefont {Goh}}]{Hu2020}%
  \BibitemOpen
  \bibfield  {author} {\bibinfo {author} {\bibfnamefont {Y.~J.}\ \bibnamefont
  {Hu}}, \bibinfo {author} {\bibfnamefont {W.~C.}\ \bibnamefont {Yu}}, \bibinfo
  {author} {\bibfnamefont {K.~T.}\ \bibnamefont {Lai}}, \bibinfo {author}
  {\bibfnamefont {D.}~\bibnamefont {Sun}}, \bibinfo {author} {\bibfnamefont
  {F.~F.}\ \bibnamefont {Balakirev}}, \bibinfo {author} {\bibfnamefont
  {W.}~\bibnamefont {Zhang}}, \bibinfo {author} {\bibfnamefont {J.~Y.}\
  \bibnamefont {Xie}}, \bibinfo {author} {\bibfnamefont {K.~Y.}\ \bibnamefont
  {Yip}}, \bibinfo {author} {\bibfnamefont {E.~I.~P.}\ \bibnamefont
  {Aulestia}}, \bibinfo {author} {\bibfnamefont {R.}~\bibnamefont {Jha}},
  \bibinfo {author} {\bibfnamefont {R.}~\bibnamefont {Higashinaka}}, \bibinfo
  {author} {\bibfnamefont {T.~D.}\ \bibnamefont {Matsuda}}, \bibinfo {author}
  {\bibfnamefont {Y.}~\bibnamefont {Yanase}}, \bibinfo {author} {\bibfnamefont
  {Y.}~\bibnamefont {Aoki}},\ and\ \bibinfo {author} {\bibfnamefont {S.~K.}\
  \bibnamefont {Goh}},\ }\bibfield  {title} {\enquote {\bibinfo {title}
  {Detection of hole pockets in the candidate type-{II Weyl} semimetal
  {${\mathrm{MoTe}}_{2}$} from {Shubnikov--de Haas} quantum oscillations},}\
  }\href {https://doi.org/10.1103/PhysRevLett.124.076402} {\bibfield  {journal}
  {\bibinfo  {journal} {Phys. Rev. Lett.}\ }\textbf {\bibinfo {volume} {124}},\
  \bibinfo {pages} {076402} (\bibinfo {year} {2020})}\BibitemShut {NoStop}%
\bibitem [{\citenamefont {Swift}\ and\ \citenamefont
  {Packard}(1979)}]{Swift1979}%
  \BibitemOpen
  \bibfield  {author} {\bibinfo {author} {\bibfnamefont {G.}~\bibnamefont
  {Swift}}\ and\ \bibinfo {author} {\bibfnamefont {R.}~\bibnamefont
  {Packard}},\ }\bibfield  {title} {\enquote {\bibinfo {title} {Thermal
  contraction of {Vespel SP-22} and {Stycast} 1266 from 300{K} to 4{K}},}\
  }\href {https://doi.org/https://doi.org/10.1016/0011-2275(79)90161-9}
  {\bibfield  {journal} {\bibinfo  {journal} {Cryogenics}\ }\textbf {\bibinfo
  {volume} {19}},\ \bibinfo {pages} {362--363} (\bibinfo {year}
  {1979})}\BibitemShut {NoStop}%
\bibitem [{\citenamefont {Werthamer}, \citenamefont {Helfand},\ and\
  \citenamefont {Hohenberg}(1966)}]{Werthamer1966}%
  \BibitemOpen
  \bibfield  {author} {\bibinfo {author} {\bibfnamefont {N.~R.}\ \bibnamefont
  {Werthamer}}, \bibinfo {author} {\bibfnamefont {E.}~\bibnamefont {Helfand}},\
  and\ \bibinfo {author} {\bibfnamefont {P.~C.}\ \bibnamefont {Hohenberg}},\
  }\bibfield  {title} {\enquote {\bibinfo {title} {Temperature and purity
  dependence of the superconducting critical field, ${H}_{c2}$. iii. electron
  spin and spin-orbit effects},}\ }\href@noop {} {\bibfield  {journal}
  {\bibinfo  {journal} {Phys. Rev.}\ }\textbf {\bibinfo {volume} {147}},\
  \bibinfo {pages} {295--302} (\bibinfo {year} {1966})}\BibitemShut {NoStop}%
\bibitem [{\citenamefont {Tinkham}(2004)}]{tinkham2004}%
  \BibitemOpen
  \bibfield  {author} {\bibinfo {author} {\bibfnamefont {M.}~\bibnamefont
  {Tinkham}},\ }\href@noop {} {\emph {\bibinfo {title} {Introduction to
  superconductivity}}}\ (\bibinfo  {publisher} {Courier Corporation},\ \bibinfo
  {year} {2004})\BibitemShut {NoStop}%
\bibitem [{\citenamefont {Chan}\ \emph {et~al.}(2017)\citenamefont {Chan},
  \citenamefont {Alireza}, \citenamefont {Yip}, \citenamefont {Niu},
  \citenamefont {Lai},\ and\ \citenamefont {Goh}}]{Chan2017}%
  \BibitemOpen
  \bibfield  {author} {\bibinfo {author} {\bibfnamefont {Y.~T.}\ \bibnamefont
  {Chan}}, \bibinfo {author} {\bibfnamefont {P.~L.}\ \bibnamefont {Alireza}},
  \bibinfo {author} {\bibfnamefont {K.~Y.}\ \bibnamefont {Yip}}, \bibinfo
  {author} {\bibfnamefont {Q.}~\bibnamefont {Niu}}, \bibinfo {author}
  {\bibfnamefont {K.~T.}\ \bibnamefont {Lai}},\ and\ \bibinfo {author}
  {\bibfnamefont {S.~K.}\ \bibnamefont {Goh}},\ }\bibfield  {title} {\enquote
  {\bibinfo {title} {Nearly isotropic superconductivity in the layered {Weyl}
  semimetal {${\mathrm{WTe}}_{2}$} at 98.5 kbar},}\ }\href
  {https://doi.org/10.1103/PhysRevB.96.180504} {\bibfield  {journal} {\bibinfo
  {journal} {Phys. Rev. B}\ }\textbf {\bibinfo {volume} {96}},\ \bibinfo
  {pages} {180504} (\bibinfo {year} {2017})}\BibitemShut {NoStop}%
\bibitem [{\citenamefont {Tinkham}(1963)}]{Tinkham1963}%
  \BibitemOpen
  \bibfield  {author} {\bibinfo {author} {\bibfnamefont {M.}~\bibnamefont
  {Tinkham}},\ }\bibfield  {title} {\enquote {\bibinfo {title} {Effect of
  fluxoid quantization on transitions of superconducting films},}\ }\href
  {https://doi.org/10.1103/PhysRev.129.2413} {\bibfield  {journal} {\bibinfo
  {journal} {Phys. Rev.}\ }\textbf {\bibinfo {volume} {129}},\ \bibinfo {pages}
  {2413--2422} (\bibinfo {year} {1963})}\BibitemShut {NoStop}%
\bibitem [{\citenamefont {Rhodes}\ \emph {et~al.}(2017)\citenamefont {Rhodes},
  \citenamefont {Sch\"onemann}, \citenamefont {Aryal}, \citenamefont {Zhou},
  \citenamefont {Zhang}, \citenamefont {Kampert}, \citenamefont {Chiu},
  \citenamefont {Lai}, \citenamefont {Shimura}, \citenamefont {McCandless},
  \citenamefont {Chan}, \citenamefont {Paley}, \citenamefont {Lee},
  \citenamefont {Finke}, \citenamefont {Ruff}, \citenamefont {Das},
  \citenamefont {Manousakis},\ and\ \citenamefont {Balicas}}]{Rhodes2017}%
  \BibitemOpen
  \bibfield  {author} {\bibinfo {author} {\bibfnamefont {D.}~\bibnamefont
  {Rhodes}}, \bibinfo {author} {\bibfnamefont {R.}~\bibnamefont
  {Sch\"onemann}}, \bibinfo {author} {\bibfnamefont {N.}~\bibnamefont {Aryal}},
  \bibinfo {author} {\bibfnamefont {Q.}~\bibnamefont {Zhou}}, \bibinfo {author}
  {\bibfnamefont {Q.~R.}\ \bibnamefont {Zhang}}, \bibinfo {author}
  {\bibfnamefont {E.}~\bibnamefont {Kampert}}, \bibinfo {author} {\bibfnamefont
  {Y.-C.}\ \bibnamefont {Chiu}}, \bibinfo {author} {\bibfnamefont
  {Y.}~\bibnamefont {Lai}}, \bibinfo {author} {\bibfnamefont {Y.}~\bibnamefont
  {Shimura}}, \bibinfo {author} {\bibfnamefont {G.~T.}\ \bibnamefont
  {McCandless}}, \bibinfo {author} {\bibfnamefont {J.~Y.}\ \bibnamefont
  {Chan}}, \bibinfo {author} {\bibfnamefont {D.~W.}\ \bibnamefont {Paley}},
  \bibinfo {author} {\bibfnamefont {J.}~\bibnamefont {Lee}}, \bibinfo {author}
  {\bibfnamefont {A.~D.}\ \bibnamefont {Finke}}, \bibinfo {author}
  {\bibfnamefont {J.~P.~C.}\ \bibnamefont {Ruff}}, \bibinfo {author}
  {\bibfnamefont {S.}~\bibnamefont {Das}}, \bibinfo {author} {\bibfnamefont
  {E.}~\bibnamefont {Manousakis}},\ and\ \bibinfo {author} {\bibfnamefont
  {L.}~\bibnamefont {Balicas}},\ }\bibfield  {title} {\enquote {\bibinfo
  {title} {Bulk {Fermi} surface of the {Weyl} type-{II} semimetallic candidate
  {$\ensuremath{\gamma}\ensuremath{-}{\mathrm{MoTe}}_{2}$}},}\ }\href
  {https://doi.org/10.1103/PhysRevB.96.165134} {\bibfield  {journal} {\bibinfo
  {journal} {Phys. Rev. B}\ }\textbf {\bibinfo {volume} {96}},\ \bibinfo
  {pages} {165134} (\bibinfo {year} {2017})}\BibitemShut {NoStop}%
\bibitem [{\citenamefont {Liu}\ \emph {et~al.}(2020)\citenamefont {Liu},
  \citenamefont {Heikes}, \citenamefont {Yildirim}, \citenamefont {Eckberg},
  \citenamefont {Metz}, \citenamefont {Kim}, \citenamefont {Ran}, \citenamefont
  {Ratcliff}, \citenamefont {Paglione},\ and\ \citenamefont {Butch}}]{Liu2020}%
  \BibitemOpen
  \bibfield  {author} {\bibinfo {author} {\bibfnamefont {I.-L.}\ \bibnamefont
  {Liu}}, \bibinfo {author} {\bibfnamefont {C.}~\bibnamefont {Heikes}},
  \bibinfo {author} {\bibfnamefont {T.}~\bibnamefont {Yildirim}}, \bibinfo
  {author} {\bibfnamefont {C.}~\bibnamefont {Eckberg}}, \bibinfo {author}
  {\bibfnamefont {T.}~\bibnamefont {Metz}}, \bibinfo {author} {\bibfnamefont
  {H.}~\bibnamefont {Kim}}, \bibinfo {author} {\bibfnamefont {S.}~\bibnamefont
  {Ran}}, \bibinfo {author} {\bibfnamefont {W.~D.}\ \bibnamefont {Ratcliff}},
  \bibinfo {author} {\bibfnamefont {J.}~\bibnamefont {Paglione}},\ and\
  \bibinfo {author} {\bibfnamefont {N.~P.}\ \bibnamefont {Butch}},\ }\bibfield
  {title} {\enquote {\bibinfo {title} {Quantum oscillations from networked
  topological interfaces in a weyl semimetal},}\ }\href
  {https://doi.org/10.1038/s41535-020-00264-8} {\bibfield  {journal} {\bibinfo
  {journal} {npj Quantum Mater.}\ }\textbf {\bibinfo {volume} {5}},\ \bibinfo
  {pages} {62} (\bibinfo {year} {2020})}\BibitemShut {NoStop}%
\bibitem [{\citenamefont {Frindt}(1972)}]{Frindt1972}%
  \BibitemOpen
  \bibfield  {author} {\bibinfo {author} {\bibfnamefont {R.~F.}\ \bibnamefont
  {Frindt}},\ }\bibfield  {title} {\enquote {\bibinfo {title}
  {Superconductivity in {U}ltrathin {NbSe$_2$} {L}ayers},}\ }\href
  {https://doi.org/10.1103/PhysRevLett.28.299} {\bibfield  {journal} {\bibinfo
  {journal} {Phys. Rev. Lett.}\ }\textbf {\bibinfo {volume} {28}},\ \bibinfo
  {pages} {299--301} (\bibinfo {year} {1972})}\BibitemShut {NoStop}%
\bibitem [{\citenamefont {Sipos}\ \emph {et~al.}(2008)\citenamefont {Sipos},
  \citenamefont {Kusmartseva}, \citenamefont {Akrap}, \citenamefont {Berger},
  \citenamefont {Forro},\ and\ \citenamefont {Tutis}}]{Sipos2008}%
  \BibitemOpen
  \bibfield  {author} {\bibinfo {author} {\bibfnamefont {B.}~\bibnamefont
  {Sipos}}, \bibinfo {author} {\bibfnamefont {A.~F.}\ \bibnamefont
  {Kusmartseva}}, \bibinfo {author} {\bibfnamefont {A.}~\bibnamefont {Akrap}},
  \bibinfo {author} {\bibfnamefont {H.}~\bibnamefont {Berger}}, \bibinfo
  {author} {\bibfnamefont {L.}~\bibnamefont {Forro}},\ and\ \bibinfo {author}
  {\bibfnamefont {E.}~\bibnamefont {Tutis}},\ }\bibfield  {title} {\enquote
  {\bibinfo {title} {From {M}ott state to superconductivity in {1T-TaS$_2$}},}\
  }\href {https://doi.org/10.1038/nmat2318} {\bibfield  {journal} {\bibinfo
  {journal} {Nat. Mater.}\ }\textbf {\bibinfo {volume} {7}},\ \bibinfo {pages}
  {960--965} (\bibinfo {year} {2008})}\BibitemShut {NoStop}%
\bibitem [{\citenamefont {Ye}\ \emph {et~al.}(2012)\citenamefont {Ye},
  \citenamefont {Zhang}, \citenamefont {Akashi}, \citenamefont {Bahramy},
  \citenamefont {Arita},\ and\ \citenamefont {Iwasa}}]{Ye2012}%
  \BibitemOpen
  \bibfield  {author} {\bibinfo {author} {\bibfnamefont {J.~T.}\ \bibnamefont
  {Ye}}, \bibinfo {author} {\bibfnamefont {Y.~J.}\ \bibnamefont {Zhang}},
  \bibinfo {author} {\bibfnamefont {R.}~\bibnamefont {Akashi}}, \bibinfo
  {author} {\bibfnamefont {M.~S.}\ \bibnamefont {Bahramy}}, \bibinfo {author}
  {\bibfnamefont {R.}~\bibnamefont {Arita}},\ and\ \bibinfo {author}
  {\bibfnamefont {Y.}~\bibnamefont {Iwasa}},\ }\bibfield  {title} {\enquote
  {\bibinfo {title} {Superconducting dome in a gate-tuned band insulator},}\
  }\href {https://doi.org/10.1126/science.1228006} {\bibfield  {journal}
  {\bibinfo  {journal} {Science}\ }\textbf {\bibinfo {volume} {338}},\ \bibinfo
  {pages} {1193--1196} (\bibinfo {year} {2012})}\BibitemShut {NoStop}%
\bibitem [{\citenamefont {Taniguchi}\ \emph {et~al.}(2012)\citenamefont
  {Taniguchi}, \citenamefont {Matsumoto}, \citenamefont {Shimotani},\ and\
  \citenamefont {Takagi}}]{Taniguchi2012}%
  \BibitemOpen
  \bibfield  {author} {\bibinfo {author} {\bibfnamefont {K.}~\bibnamefont
  {Taniguchi}}, \bibinfo {author} {\bibfnamefont {A.}~\bibnamefont
  {Matsumoto}}, \bibinfo {author} {\bibfnamefont {H.}~\bibnamefont
  {Shimotani}},\ and\ \bibinfo {author} {\bibfnamefont {H.}~\bibnamefont
  {Takagi}},\ }\bibfield  {title} {\enquote {\bibinfo {title}
  {Electric-field-induced superconductivity at 9.4 {K} in a layered transition
  metal disulphide {MoS$_2$}},}\ }\href@noop {} {\bibfield  {journal} {\bibinfo
   {journal} {Appl. Phys. Lett.}\ }\textbf {\bibinfo {volume} {101}} (\bibinfo
  {year} {2012})}\BibitemShut {NoStop}%
\bibitem [{\citenamefont {Kang}\ \emph {et~al.}(2015)\citenamefont {Kang},
  \citenamefont {Zhou}, \citenamefont {Yi}, \citenamefont {Yang}, \citenamefont
  {Guo}, \citenamefont {Shi}, \citenamefont {Zhang}, \citenamefont {Wang},
  \citenamefont {Zhang}, \citenamefont {Jiang}, \citenamefont {Li},
  \citenamefont {Yang}, \citenamefont {Wu}, \citenamefont {Zhang},
  \citenamefont {Sun},\ and\ \citenamefont {Zhao}}]{Kang2015}%
  \BibitemOpen
  \bibfield  {author} {\bibinfo {author} {\bibfnamefont {D.}~\bibnamefont
  {Kang}}, \bibinfo {author} {\bibfnamefont {Y.}~\bibnamefont {Zhou}}, \bibinfo
  {author} {\bibfnamefont {W.}~\bibnamefont {Yi}}, \bibinfo {author}
  {\bibfnamefont {C.}~\bibnamefont {Yang}}, \bibinfo {author} {\bibfnamefont
  {J.}~\bibnamefont {Guo}}, \bibinfo {author} {\bibfnamefont {Y.}~\bibnamefont
  {Shi}}, \bibinfo {author} {\bibfnamefont {S.}~\bibnamefont {Zhang}}, \bibinfo
  {author} {\bibfnamefont {Z.}~\bibnamefont {Wang}}, \bibinfo {author}
  {\bibfnamefont {C.}~\bibnamefont {Zhang}}, \bibinfo {author} {\bibfnamefont
  {S.}~\bibnamefont {Jiang}}, \bibinfo {author} {\bibfnamefont
  {A.}~\bibnamefont {Li}}, \bibinfo {author} {\bibfnamefont {K.}~\bibnamefont
  {Yang}}, \bibinfo {author} {\bibfnamefont {Q.}~\bibnamefont {Wu}}, \bibinfo
  {author} {\bibfnamefont {G.}~\bibnamefont {Zhang}}, \bibinfo {author}
  {\bibfnamefont {L.}~\bibnamefont {Sun}},\ and\ \bibinfo {author}
  {\bibfnamefont {Z.}~\bibnamefont {Zhao}},\ }\bibfield  {title} {\enquote
  {\bibinfo {title} {Superconductivity emerging from a suppressed large
  magnetoresistant state in tungsten ditelluride},}\ }\href@noop {} {\bibfield
  {journal} {\bibinfo  {journal} {Nat. Commun.}\ }\textbf {\bibinfo {volume}
  {6}},\ \bibinfo {pages} {7804} (\bibinfo {year} {2015})}\BibitemShut
  {NoStop}%
\bibitem [{\citenamefont {Pan}\ \emph {et~al.}(2015)\citenamefont {Pan},
  \citenamefont {Chen}, \citenamefont {Liu}, \citenamefont {Feng},
  \citenamefont {Wei}, \citenamefont {Zhou}, \citenamefont {Chi}, \citenamefont
  {Pi}, \citenamefont {Yen}, \citenamefont {Song}, \citenamefont {Wan},
  \citenamefont {Yang}, \citenamefont {Wang}, \citenamefont {Wang},\ and\
  \citenamefont {Zhang}}]{Pan2015}%
  \BibitemOpen
  \bibfield  {author} {\bibinfo {author} {\bibfnamefont {X.-C.}\ \bibnamefont
  {Pan}}, \bibinfo {author} {\bibfnamefont {X.}~\bibnamefont {Chen}}, \bibinfo
  {author} {\bibfnamefont {H.}~\bibnamefont {Liu}}, \bibinfo {author}
  {\bibfnamefont {Y.}~\bibnamefont {Feng}}, \bibinfo {author} {\bibfnamefont
  {Z.}~\bibnamefont {Wei}}, \bibinfo {author} {\bibfnamefont {Y.}~\bibnamefont
  {Zhou}}, \bibinfo {author} {\bibfnamefont {Z.}~\bibnamefont {Chi}}, \bibinfo
  {author} {\bibfnamefont {L.}~\bibnamefont {Pi}}, \bibinfo {author}
  {\bibfnamefont {F.}~\bibnamefont {Yen}}, \bibinfo {author} {\bibfnamefont
  {F.}~\bibnamefont {Song}}, \bibinfo {author} {\bibfnamefont {X.}~\bibnamefont
  {Wan}}, \bibinfo {author} {\bibfnamefont {Z.}~\bibnamefont {Yang}}, \bibinfo
  {author} {\bibfnamefont {B.}~\bibnamefont {Wang}}, \bibinfo {author}
  {\bibfnamefont {G.}~\bibnamefont {Wang}},\ and\ \bibinfo {author}
  {\bibfnamefont {Y.}~\bibnamefont {Zhang}},\ }\bibfield  {title} {\enquote
  {\bibinfo {title} {Pressure-driven dome-shaped superconductivity and
  electronic structural evolution in tungsten ditelluride},}\ }\href@noop {}
  {\bibfield  {journal} {\bibinfo  {journal} {Nat. Commun.}\ }\textbf {\bibinfo
  {volume} {6}},\ \bibinfo {pages} {7805} (\bibinfo {year} {2015})}\BibitemShut
  {NoStop}%
\bibitem [{\citenamefont {Kamihara}\ \emph {et~al.}(2008)\citenamefont
  {Kamihara}, \citenamefont {Watanabe}, \citenamefont {Hirano},\ and\
  \citenamefont {Hosono}}]{Kamihara2008}%
  \BibitemOpen
  \bibfield  {author} {\bibinfo {author} {\bibfnamefont {Y.}~\bibnamefont
  {Kamihara}}, \bibinfo {author} {\bibfnamefont {T.}~\bibnamefont {Watanabe}},
  \bibinfo {author} {\bibfnamefont {M.}~\bibnamefont {Hirano}},\ and\ \bibinfo
  {author} {\bibfnamefont {H.}~\bibnamefont {Hosono}},\ }\bibfield  {title}
  {\enquote {\bibinfo {title} {Iron-{B}ased {L}ayered {S}uperconductor
  {La[O$_{1-x}$F$_x$]FeAs} ($x = 0.05-0.12$) with {$T_c$} = 26 {K}},}\ }\href
  {https://doi.org/10.1021/ja800073m} {\bibfield  {journal} {\bibinfo
  {journal} {J. Am. Chem. Soc.}\ }\textbf {\bibinfo {volume} {130}},\ \bibinfo
  {pages} {3296--3297} (\bibinfo {year} {2008})}\BibitemShut {NoStop}%
\bibitem [{\citenamefont {Paglione}\ and\ \citenamefont
  {Greene}(2010)}]{Paglione2010}%
  \BibitemOpen
  \bibfield  {author} {\bibinfo {author} {\bibfnamefont {J.}~\bibnamefont
  {Paglione}}\ and\ \bibinfo {author} {\bibfnamefont {R.~L.}\ \bibnamefont
  {Greene}},\ }\bibfield  {title} {\enquote {\bibinfo {title} {High-temperature
  superconductivity in iron-based materials},}\ }\href
  {https://doi.org/10.1038/NPHYS1759} {\bibfield  {journal} {\bibinfo
  {journal} {Nat. Phys.}\ }\textbf {\bibinfo {volume} {6}},\ \bibinfo {pages}
  {645--658} (\bibinfo {year} {2010})}\BibitemShut {NoStop}%
\bibitem [{\citenamefont {Hosono}\ and\ \citenamefont
  {Kuroki}(2015)}]{Hosono2015}%
  \BibitemOpen
  \bibfield  {author} {\bibinfo {author} {\bibfnamefont {H.}~\bibnamefont
  {Hosono}}\ and\ \bibinfo {author} {\bibfnamefont {K.}~\bibnamefont
  {Kuroki}},\ }\bibfield  {title} {\enquote {\bibinfo {title} {Iron-based
  superconductors: {C}urrent status of materials and pairing mechanism},}\
  }\href {https://doi.org/https://doi.org/10.1016/j.physc.2015.02.020}
  {\bibfield  {journal} {\bibinfo  {journal} {Phys. C. Supercond.}\ }\textbf
  {\bibinfo {volume} {514}},\ \bibinfo {pages} {399--422} (\bibinfo {year}
  {2015})}\BibitemShut {NoStop}%
\bibitem [{\citenamefont {Ortiz}\ \emph {et~al.}(2019)\citenamefont {Ortiz},
  \citenamefont {Gomes}, \citenamefont {Morey}, \citenamefont {Winiarski},
  \citenamefont {Bordelon}, \citenamefont {Mangum}, \citenamefont {Oswald},
  \citenamefont {Rodriguez-Rivera}, \citenamefont {Neilson}, \citenamefont
  {Wilson}, \citenamefont {Ertekin}, \citenamefont {McQueen},\ and\
  \citenamefont {Toberer}}]{Ortiz2019}%
  \BibitemOpen
  \bibfield  {author} {\bibinfo {author} {\bibfnamefont {B.~R.}\ \bibnamefont
  {Ortiz}}, \bibinfo {author} {\bibfnamefont {L.~C.}\ \bibnamefont {Gomes}},
  \bibinfo {author} {\bibfnamefont {J.~R.}\ \bibnamefont {Morey}}, \bibinfo
  {author} {\bibfnamefont {M.}~\bibnamefont {Winiarski}}, \bibinfo {author}
  {\bibfnamefont {M.}~\bibnamefont {Bordelon}}, \bibinfo {author}
  {\bibfnamefont {J.~S.}\ \bibnamefont {Mangum}}, \bibinfo {author}
  {\bibfnamefont {I.~W.~H.}\ \bibnamefont {Oswald}}, \bibinfo {author}
  {\bibfnamefont {J.~A.}\ \bibnamefont {Rodriguez-Rivera}}, \bibinfo {author}
  {\bibfnamefont {J.~R.}\ \bibnamefont {Neilson}}, \bibinfo {author}
  {\bibfnamefont {S.~D.}\ \bibnamefont {Wilson}}, \bibinfo {author}
  {\bibfnamefont {E.}~\bibnamefont {Ertekin}}, \bibinfo {author} {\bibfnamefont
  {T.~M.}\ \bibnamefont {McQueen}},\ and\ \bibinfo {author} {\bibfnamefont
  {E.~S.}\ \bibnamefont {Toberer}},\ }\bibfield  {title} {\enquote {\bibinfo
  {title} {New kagome prototype materials: discovery of {KV$_3$Sb$_5$},
  {RbV$_3$Sb$_5$}, and {CsV$_3$Sb$_5$}},}\ }\href
  {https://doi.org/10.1103/PhysRevMaterials.3.094407} {\bibfield  {journal}
  {\bibinfo  {journal} {Phys. Rev. Mater.}\ }\textbf {\bibinfo {volume} {3}},\
  \bibinfo {pages} {094407} (\bibinfo {year} {2019})}\BibitemShut {NoStop}%
\bibitem [{\citenamefont {Ortiz}\ \emph {et~al.}(2020)\citenamefont {Ortiz},
  \citenamefont {Teicher}, \citenamefont {Hu}, \citenamefont {Zuo},
  \citenamefont {Sarte}, \citenamefont {Schueller}, \citenamefont {Abeykoon},
  \citenamefont {Krogstad}, \citenamefont {Rosenkranz}, \citenamefont {Osborn},
  \citenamefont {Seshadri}, \citenamefont {Balents}, \citenamefont {He},\ and\
  \citenamefont {Wilson}}]{Ortiz2020}%
  \BibitemOpen
  \bibfield  {author} {\bibinfo {author} {\bibfnamefont {B.~R.}\ \bibnamefont
  {Ortiz}}, \bibinfo {author} {\bibfnamefont {S.~M.~L.}\ \bibnamefont
  {Teicher}}, \bibinfo {author} {\bibfnamefont {Y.}~\bibnamefont {Hu}},
  \bibinfo {author} {\bibfnamefont {J.~L.}\ \bibnamefont {Zuo}}, \bibinfo
  {author} {\bibfnamefont {P.~M.}\ \bibnamefont {Sarte}}, \bibinfo {author}
  {\bibfnamefont {E.~C.}\ \bibnamefont {Schueller}}, \bibinfo {author}
  {\bibfnamefont {A.~M.~M.}\ \bibnamefont {Abeykoon}}, \bibinfo {author}
  {\bibfnamefont {M.~J.}\ \bibnamefont {Krogstad}}, \bibinfo {author}
  {\bibfnamefont {S.}~\bibnamefont {Rosenkranz}}, \bibinfo {author}
  {\bibfnamefont {R.}~\bibnamefont {Osborn}}, \bibinfo {author} {\bibfnamefont
  {R.}~\bibnamefont {Seshadri}}, \bibinfo {author} {\bibfnamefont
  {L.}~\bibnamefont {Balents}}, \bibinfo {author} {\bibfnamefont
  {J.}~\bibnamefont {He}},\ and\ \bibinfo {author} {\bibfnamefont {S.~D.}\
  \bibnamefont {Wilson}},\ }\bibfield  {title} {\enquote {\bibinfo {title}
  {{CsV$_3$Sb$_5$}: A {${\mathbb{Z}}_{2}$} topological kagome metal with a
  superconducting ground state},}\ }\href
  {https://doi.org/10.1103/PhysRevLett.125.247002} {\bibfield  {journal}
  {\bibinfo  {journal} {Phys. Rev. Lett.}\ }\textbf {\bibinfo {volume} {125}},\
  \bibinfo {pages} {247002} (\bibinfo {year} {2020})}\BibitemShut {NoStop}%
\bibitem [{\citenamefont {Neupert}\ \emph {et~al.}(2021)\citenamefont
  {Neupert}, \citenamefont {Denner}, \citenamefont {Yin}, \citenamefont
  {Thomale},\ and\ \citenamefont {Hasan}}]{Neupert2021}%
  \BibitemOpen
  \bibfield  {author} {\bibinfo {author} {\bibfnamefont {T.}~\bibnamefont
  {Neupert}}, \bibinfo {author} {\bibfnamefont {M.~M.}\ \bibnamefont {Denner}},
  \bibinfo {author} {\bibfnamefont {J.-X.}\ \bibnamefont {Yin}}, \bibinfo
  {author} {\bibfnamefont {R.}~\bibnamefont {Thomale}},\ and\ \bibinfo {author}
  {\bibfnamefont {M.~Z.}\ \bibnamefont {Hasan}},\ }\bibfield  {title} {\enquote
  {\bibinfo {title} {Charge order and superconductivity in kagome materials},}\
  }\href {https://doi.org/10.1038/s41567-021-01404-y} {\bibfield  {journal}
  {\bibinfo  {journal} {Nat. Phys.}\ }\textbf {\bibinfo {volume} {18}},\
  \bibinfo {pages} {137--143} (\bibinfo {year} {2021})}\BibitemShut {NoStop}%
\end{thebibliography}
\providecommand{\noopsort}[1]{}\providecommand{\singleletter}[1]{#1}%
\end{document}